\documentclass[twoside,leqno,twocolumn]{article}

\usepackage[letterpaper]{geometry}

\usepackage{ltexpprt}
\usepackage{algorithm}
\usepackage{algpseudocode}
\usepackage{amsmath,amssymb,amsfonts}
\usepackage{graphicx}
\usepackage{textcomp}
\usepackage{xcolor}
\usepackage{booktabs} 
\usepackage{amsmath}
\usepackage{graphicx}
\usepackage{subfigure}
\usepackage{multirow}
\usepackage{amsfonts,bm}
\usepackage{braket}
\usepackage{mathtools}
\usepackage{enumerate}
\usepackage{enumitem}
\usepackage{empheq}
\usepackage{calc}
\usepackage{balance}
\usepackage{dsfont}
\usepackage{textcomp} 
\usepackage{caption}
\usepackage{color, colortbl}
\usepackage{url}

\begin{document}



\title{ \Large Modeling Continuous Spatial-temporal Dynamics of Turbulent Flow with Test-time Refinement}

\author{Shengyu Chen,  Peyman Givi, Can Zheng, Xiaowei Jia\\
\small\baselineskip=9pt University of Pittsburgh\\ 
\small  \{shc160, peg10, caz51, xiaowei\}@pitt.edu
}


\date{}

\maketitle


\fancyfoot[R]{\scriptsize{Copyright \textcopyright\ 2025 by SIAM\\
Unauthorized reproduction of this article is prohibited}}





\begin{abstract}
The precise simulation of turbulent flows holds immense significance across various scientific and engineering domains, including climate science, freshwater science, and energy-efficient manufacturing. Within the realm of simulating turbulent flows, large eddy simulation (LES) has emerged as a prevalent alternative to direct numerical simulation (DNS), offering computational efficiency. However, LES cannot accurately capture the full spectrum of turbulent transport scales and is present only at a lower spatial resolution. Reconstructing high-fidelity DNS data from the lower-resolution LES data is essential for numerous applications, but it poses significant challenges to existing super-resolution techniques, primarily due to the complex spatio-temporal nature of turbulent flows.  This paper proposes a novel flow reconstruction approach that leverages physical knowledge to model flow dynamics. Different from traditional super-resolution techniques, the proposed approach uses LES data only in the testing phase through a degradation-based refinement approach to enforce physical constraints and mitigate cumulative reconstruction errors over time. Furthermore, a feature sampling strategy is developed to enable flow data reconstruction across different resolutions. The results on two distinct sets of turbulent flow data indicate the effectiveness of the proposed method in reconstructing high-resolution DNS data, preserving the inherent physical attributes of flow transport, and achieving DNS reconstruction at different resolutions.

\end{abstract}

\section{Introduction}

Advances in computational fluid dynamics (CFD) have significantly impacted many scientific and engineering domains. In the clean energy sector, CFD is crucial for enhancing power generation and distribution, which includes designing high-efficiency wind turbines and strategically placing turbines to maximize energy harvest. CFD is also important for developing safer and more efficient cooling systems used in thermal and nuclear power plants. In the aerospace industry, CFD is used to analyze aerodynamic forces and thermal effects on aircraft, rockets, and spacecraft. It plays a vital role in simulating and optimizing airflow around wings, fuselages, and engine components. The optimization helps improve fuel efficiency, reduce drag, enhance maneuverability, and increase safety. 
Additionally, CFD is critically needed in many environmental science problems, 
e.g.,  predicting pollution patterns, refining emission controls, and assessing the environmental impact of infrastructure. It is especially important for studying ocean currents and their impact on climate change. By simulating complex ocean dynamics, including factors such as temperature and salinity and their interactions with atmospheric conditions,  CFD helps scientists understand and predict changes in major oceanic systems, such as the Gulf Stream~\cite{taylor1998gulf} or the El Niño phenomenon~\cite{diez2005nino}. These simulations are vital for predicting climate change effects, assessing impacts on marine ecosystems, and informing policy decisions related to climate change mitigation and adaptation~\cite{qu2019integration}.

In particular, turbulent flows often need to be simulated at high spatial resolutions and over long periods in many CFD tasks (e.g.,  simulating fine-level cloud behaviors for climate models~\cite{rasp2018deep}). 
Direct Numerical Simulation (DNS) has been widely considered as the method with the highest fidelity in creating turbulence simulations, but 
it is computationally intensive and thus limited in producing long-term simulations at fine spatial scales~\cite{Givi94}. 
As an alternative with reduced computational cost, large eddy simulation (LES) has gained popularity. LES focuses on the larger scale energy-containing eddies while filtering out the smaller scales of transport~\cite{Sagaut05}. Consequently, LES can be performed on coarser grids compared to DNS, but at the expense of reduced fidelity~\cite{NNGLP17}.

Machine learning methods, including super-resolution (SR) techniques~\cite{Cheo2003SR}, have been advocated as a solution for reconstructing highly detailed DNS from LES data. These approaches have demonstrated remarkable success in enhancing high-resolution data across various commercial applications. Most contemporary SR models utilize convolutional network layers (CNNs)~\cite{albawi2017understanding} to extract meaningful spatial features, which are then used to recover high-resolution images through non-linear transformation. From the initial end-to-end convolutional SRCNN model~\cite{dong2014learning}, numerous researchers have incorporated additional structural elements, including skip-connections~\cite{zhang2018residual,ahn2018fast,Dai2019,Duong2021}, channel attention~\cite{zhang2018image}, adversarial training objectives~\cite{ledig2017photo,wang2018recovering,wang2018esrgan,karras2018progressive,gan8759375,cheng2021mfagan,Long2021}, implicit neural representation~\cite{skorokhodov2021adversarial,chen2021learning,boussif2022magnet,chen2022videoinr}, and more recently, Transformer-based SR structures~\cite{fang2022cross,lu2022transformer,fang2022hybrid,wang2022detail,zou2022self,liang2022light}. 

Given their prominence in the field of computer vision, SR techniques are gaining popularity for reconstructing high-resolution turbulence data (e.g., DNS) from low-resolution turbulence data (e.g., LES) of lower fidelity~\cite{fukami2019super,Fukami_2020,liu2020deep,Deng2019SuperresolutionRO,yang2023super,xu2023super}. However, these techniques remain limited in recovering fine-level flow patterns 
due to the absence of physical information (i.e., small-scale flow transport) in the input low-resolution data. In an effort to preserve fine-scale physical patterns and capture the temporal dependencies in simulating turbulence data, an alternative sequential prediction method has been explored 
to simulate high-resolution flow data from historical high-resolution data without referring to low-resolution data. These approaches utilize temporal modeling structures to capture underlying dynamics expressed by governing partial differential equations (PDEs) i.e., the Navier-Stokes equation,  through the utilization of neural operators~\cite{li2020fourier,equer2023multi,boussif2022magnet}, or through the recurrent unit that directly encodes the Navier-Stokes equation~\cite{bao2022physics}. Although these methods can better preserve fine-level physical patterns from input high-resolution data, they still rely on data-driven learning structures to simulate complex temporal dynamics and can be susceptible to cumulative errors over time. Furthermore, existing SR and sequential methods for turbulent flows often assume a predefined grid structure and lack the ability to simulate flow variables at arbitrary positions and at different scales. These models need to be re-trained to accommodate a 
target grid structure or spatial resolution that differs from the historical data.  

To address these challenges, a novel physics-guided neural network named "\textbf{S}uper-\textbf{R}esolution through \textbf{T}est-time \textbf{R}efinement" (SR-TR) has been developed. The SR-TR method is composed of two main components: the degradation-based refinement and the continuous spatial transition unit (CSTU). 
Unlike conventional SR techniques that directly reconstruct high-resolution data from low-resolution data, SR-TR adopts the sequential prediction approach to better preserve fine-level patterns in predicting the high-resolution data while utilizing the low-resolution flow data to reduce the accumulated errors in the testing phase. In particular, SR-TR automatically adjusts the reconstructed high-resolution data in the testing phase by preserving the consistency with the available low-resolution data and known physical constraints. The CSTU structure is further designed to better capture the continuous spatial and temporal dynamics of turbulent flows during the sequential prediction process.  The CSTU structure captures the fluid dynamics by using the physics-guided recurrent unit (PRU)~\cite{bao2022physics}, which is well-suited for modeling the behavior of turbulent flows driven by complex PDEs. Additionally, CSTU facilitates the reconstruction of flow data at arbitrary locations and scales by integrating the implicit neural representation (INR) method. This method enables the CSTU to operate effectively across different resolutions, adding versatility to the SR-TR method.

\begin{figure} [!t] 
\centering
\includegraphics[width=0.8\columnwidth]{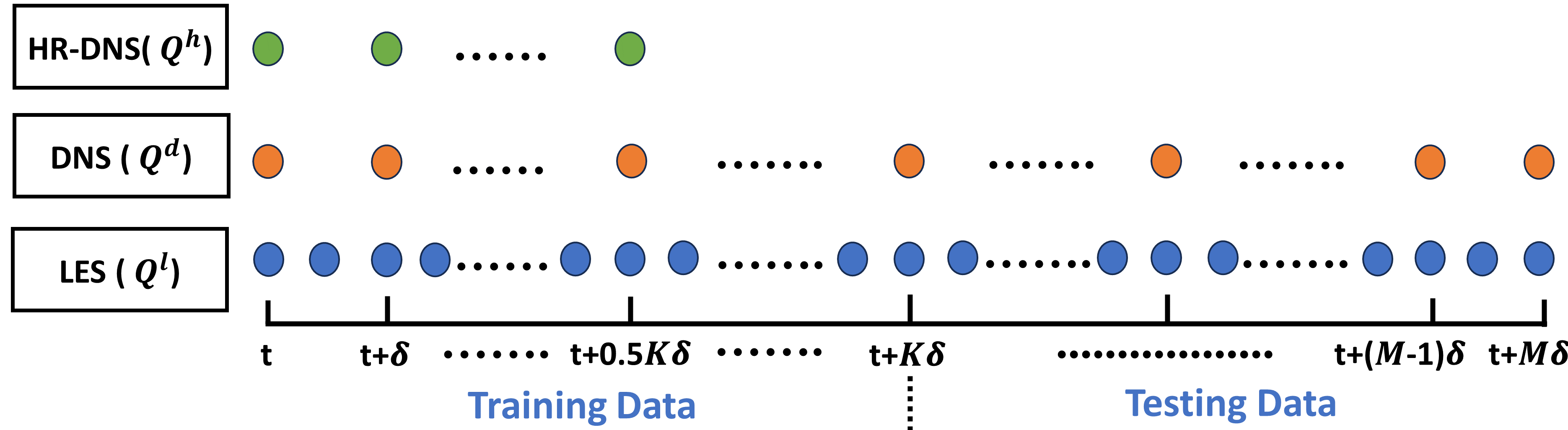}
\vspace{-0.1in}
\caption{The data distribution of LES, DNS, and HR-DNS. HR-DNS denotes the higher-resolution DNS . }
\label{fig:distribution}
\end{figure}
Comprehensive evaluations of the SR-TR method have been conducted on two distinct turbulent flow datasets: (1)~the forced isotropic turbulent (FIT) flow~\cite{FITdata}, and (2) the Taylor-Green vortex (TGV) flow~\cite{brachet1984taylor}. The results of the consistency assessments demonstrate the capability of the SR-TR capability in terms of reconstruction performance over time and across different resolutions. Furthermore, the effectiveness of each component of the methodology is demonstrated both qualitatively and quantitatively. Our implementation is publicly available$\footnote{\url{https://drive.google.com/drive/folders/109QSpDUKa3T__9Hdar9w3Pl1iHXCZZdg?usp=sharing}}$.

\section{Problem Definition and Preliminaries}
This study focuses on simulating the transport of unsteady, three-dimensional turbulent flows. In all cases, the flow is considered to be Newtonian and incompressible, with a constant density. In this formulation, the spatial coordinates are represented by the vector ${\bf x} \equiv {x, y, z}$, and time is denoted as $t$. The velocity field is denoted by ${\bf Q} ({\bf x}, t)$, with its three components ${u({\bf x}, t), v({\bf x}, t), w({\bf x}, t)}$ along the three flow directions ${x, y, z}$, respectively. The pressure, density, and dynamic viscosity are represented by $p({\bf x}, t)$, $\rho({\bf x}, t)$, and $\nu$, respectively. 
During the training process, the provided DNS data are available at a regular time interval $\delta$, denoted as $\textbf{Q}^d =\{\textbf{Q}^d(t)\}$, where $t$ belongs to the time range $\{t_0,t_0+\delta,\dots, t_0+K\delta\}$. The objective is to predict high-resolution DNS data following the provided historical data, specifically at time instances $\{t_0\!+\!(K\!+\!1)\delta, \dots, t_0\!+\!M\delta\}$.  
We also have access to low-resolution LES data,  
$\textbf{Q}^l = \{\textbf{Q}^l(t)\}$ for $t\in [t_0,t_0+M\delta]$. As the LES data can be generated at a lower computational cost, they are available for both the training and testing periods and are often generated at a higher frequency. In addition to the LES and DNS data, we assume the access to a small number of DNS data samples at an even higher resolution than $\textbf{Q}^d$, denoted as $\textbf{Q}^h = \{\textbf{Q}^h(t)\}$. These data samples are optional, and they are used only for enhancing the performance of turbulence reconstruction at resolutions higher than $\textbf{Q}^d$. Here we consider such higher-resolution DNS (HR-DNS) is available only for a shorter period, within the time range $\{t_0,t_0\!+\!\delta,\dots, t_0\!+0.5K\delta\}$ during the training period due to the high cost of generating such data.
The available data are illustrated in 
 Fig.~\ref{fig:distribution}.

\begin{figure} [!t] 
\vspace{-.1in}
\centering
\includegraphics[width=0.65\columnwidth]{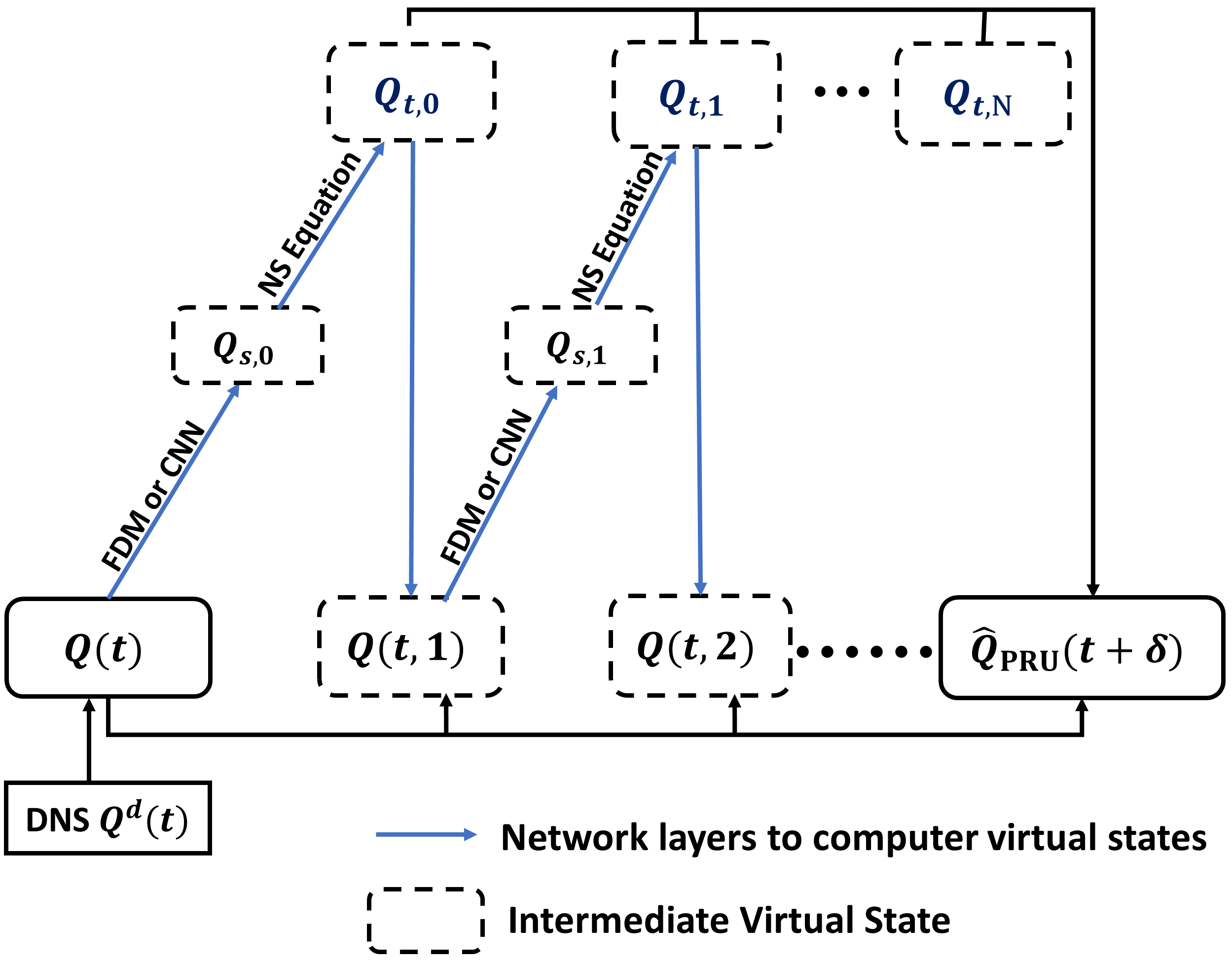}
\vspace{-0.1in}
\caption{The PRU, based on the Navier-Stokes equation, reconstructs turbulent flow data in the spatio-temporal field, where $\textbf{Q}_{s,n}$ and $\textbf{Q}_{t,n}$ are the spatial and temporal derivatives at each intermediate time step $n$.}
\label{fig:PRU}
\end{figure}

\begin{figure*} [!t] 
\vspace{-.2in}
\centering
\includegraphics[width=0.7\textwidth
]{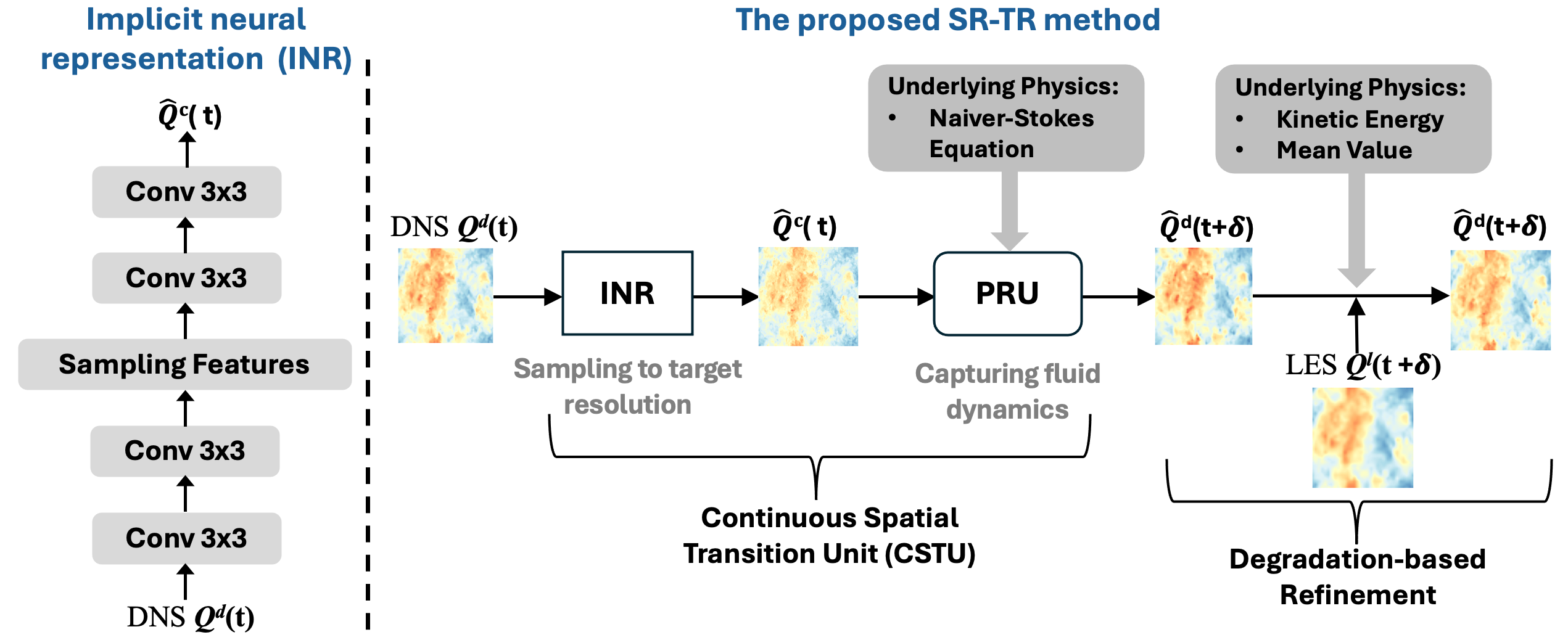}
\vspace{-0.1in}
\caption{The overall structure of SR-TR, which comprises the degradation-based refinement and the CSTU. 
}
\vspace{-0.2in}
\label{fig:Structure}
\end{figure*}

\subsection{Physics-guided Recurrent Unit (PRU)}
The physics-guided recurrent unit
(PRU)  method~\cite{bao2022physics} is built upon the Runge–Kutta (RK) discretization method~\cite{butcher2007runge}. The central idea is to utilize the continuous physical relationship represented by the underlying PDE in order to connect the discrete data samples with the continuous flow dynamics. This approach is adaptable to various dynamical systems governed by deterministic PDEs. The PDE for the target variables $\textbf{Q}$ can be formulated as: $\textbf{Q}_t = {\textbf{f}}(t, \textbf{Q};\theta)$,
where $\textbf{Q}_t$ denotes the temporal derivative of $\textbf{Q}$, and ${\textbf{f}}(t, \textbf{Q};\theta)$ is a non-linear function parameterized by the coefficient $\theta$, which summarizes the current value of $\textbf{Q}$ and its spatial variations. The turbulent data adhere to the Navier-Stokes equation for an incompressible flow:
\begin{equation}
\small
{\textbf{f}(\textbf{Q})} = -\frac{1}{\rho} \nabla p + \nu \Delta \textbf{Q} - (\textbf{Q}\cdot \nabla) \textbf{Q},
\label{eq:NS}
\end{equation} 
where $\nabla$ represents the gradient operator, and $\Delta = \nabla\cdot\nabla$ applies to each of the velocity components. 
Figure \ref{fig:PRU} illustrates the PRU structure, involving a series of intermediate states $\{\textbf{Q}(t,0),\textbf{Q}(t,1),\textbf{Q}(t,2),\dots,\textbf{Q}(t,N)\}$. The temporal gradients are estimated at these states as $\{\textbf{Q}_{t,0},\textbf{Q}_{t,1},\textbf{Q}_{t,2},\dots,\textbf{Q}_{t,N}\}$. Beginning with $\textbf{Q}(t,0)=\textbf{Q}(t)$ (where $\textbf{Q}(t)$ corresponds to $\textbf{Q}^d(t)$), the PRU method approximates the temporal gradient as $\textbf{Q}_{t,0}$ and then adjusts $\textbf{Q}(t)$ in the direction of the gradient to generate the subsequent intermediate state $\textbf{Q}(t,1)$. This process repeats for $N$ intermediate states. For the fourth-order Runge-Kutta method utilized in this work, $N=3$. Finally, PRU combines all the intermediate temporal derivatives into a composite gradient to calculate the final prediction of the next step flow data $\hat{\textbf{Q}}_{\text{PRU}}(t+\delta)$, as $\hat{\textbf{Q}}_{\text{PRU}}(t+\delta)= \textbf{Q}(t) + \sum_{n=0}^N w_n \textbf{Q}_{t,n}$, where $\{w_n\}_{n=1}^N$ are the trainable model parameters.

In more detail, the PRU method estimates temporal derivatives using the function ${\textbf{f}}(\cdot)$. According to Eq.(\ref{eq:NS}), evaluating ${\textbf{f}}(\cdot)$ requires the explicit estimation of first-order and second-order spatial derivatives. The most common approaches for estimating spatial derivatives involve finite difference methods (FDMs)~\cite{thomas2013numerical} or convolutional neural network layers (CNNs)~\cite{bao2022physics}. After estimating the first-order and second-order spatial derivatives, they are incorporated into Eq.(\ref{eq:NS}) to derive the temporal derivative $\textbf{Q}_{t,n}$.

\section{Proposed Method}
The proposed SR-TR method comprises two primary components: the degradation-based refinement and the continuous spatial transition unit (CSTU). The objective of the degradation-based refinement is to reduce the accumulated errors made by the sequential prediction by aligning the reconstructed data with low-resolution LES data while ensuring the adherence to physical constraints. In addition, the CSTU is constructed 
to capture the spatial and temporal dynamics of turbulent flows and 
facilitate the reconstruction of flow data at different resolutions. 

Figure \ref{fig:Structure} provides an overall structure of the SR-TR method. Specifically, the sequential prediction process initially occurs within the CSTU. During this process, the input DNS data $\textbf{Q}^d(t)$ at time $t$ is encoded and interpolated through the implicit neural representation method~\cite{chen2021learning,chen2022videoinr} to the desired target resolution of the flow data denoted as $\hat{\textbf{Q}}^\text{c}(t)$. Subsequently, $\hat{\textbf{Q}}^\text{c}(t)$ is fed into PRU to predict the DNS output $\hat{\textbf{Q}}^d(t+\delta)$ at the next time $t+\delta$. Degradation-based refinement is introduced during the testing phase to improve prediction and ensure consistency with physical constraints. 

\subsection{Degradation-based Refinement}

The proposed SR-TR method uses the CSTU to conduct sequential prediction from the data at time $t$ to predict  $\hat{\textbf{Q}}^d(t+\delta)$ (more details will be described later). 
With the true DNS data $\textbf{Q}^d(t+\delta)$ available in the training set, the reconstruction loss $\mathcal{L}_\text{recon}$ is defined using the mean squared error (MSE) loss, 
as $\mathcal{L}_{\text{recon}} = \sum_t\text{MSE}(\hat{\textbf{Q}}^d(t+\delta),\textbf{Q}^d(t+\delta))$.

After that, the degradation-based refinement is introduced during the testing phase. The refinement's objective is to alleviate the accumulated errors and structural distortions that arise during long-term predictions by enforcing physical consistency. 
The refinement process involves the direct downsampling of reconstructed DNS data  $\hat{\textbf{Q}}^d$ to the corresponding low-resolution LES data $\hat{\textbf{Q}}^l$. The degradation loss $\mathcal{L}_{\text{deg}}$ between $\hat{\textbf{Q}}^l$ and real LES data $\textbf{Q}^l$ is described as:
\vspace{-.1in}
\begin{equation}
\small
    \mathcal{L}_{\text{deg}} = \text{MSE}(\hat{\textbf{Q}}^l,\textbf{Q}^l).
    \label{eq:deg}
\vspace{-.1in}
\end{equation}

Another reason for utilizing  LES data in degradation loss $\mathcal{L}_{\text{deg}}$ is the unavailability of  DNS data during the testing phase. Hence, it is not feasible to directly minimize the difference between true DNS $\textbf{Q}^d$ and the reconstructed $\hat{\textbf{Q}}^d$.

Additionally, two physical constraints are introduced to ensure the consistency of (i) the mean velocity field, and (ii)~the kinetic energy of turbulence. For (i), as mean values from the true DNS data cannot be accessed, the mean values from the LES data are employed as an approximation. The equal-mean loss $\mathcal{L}_{\text{mean}}$ is defined as the difference between the mean value of reconstructed flow data $\hat{\textbf{Q}}^d$ and that of the true LES data, as follows: 
\vspace{-.05in}
\begin{equation}
\small
    \mathcal{L}_{\text{mean}} = |\overline{\textbf{Q}^l} -\overline{\hat{\textbf{Q}}^d}|.
    \label{eq:mean}
\vspace{-.05in}
\end{equation}
For (ii),  the kinetic energy is defined using three components of $\textbf{Q}$ as $\mathcal{K} = \frac{1}{2}(u^2 + v^2 + w^2)$. 
Similarly, the exact kinetic energy of the flow data is not available during the testing period. However, 
the value of kinetic energy for incompressible flows often follows simple patterns, e.g., 
constant or linearly decayed, and thus can be approximated from the DNS data in the training period, denoted by $\tilde{\mathcal{K}}$. 
Therefore, the loss function $\mathcal{L}_{\text{kinetic}}$ is defined as follows:
\vspace{-.05in}
\begin{equation}
\small
    \mathcal{L}_{\text{kinetic}} = |\mathcal{K}(\hat{\textbf{Q}}^d)-\tilde{\mathcal{K}}|.
    \label{eq:kinetic}
\vspace{-.05in}
\end{equation}

The final refinement loss function 
combines the degradation loss and physical consistency, as follows: 
\vspace{-.1in}
\begin{equation}
\small
\mathcal{L}_{\text{refine}} = \alpha_{0} \mathcal{L}_{\text{deg}} + \alpha_{1} \mathcal{L}_{\text{mean}}+
    \alpha_{2} \mathcal{L}_{\text{kinetic}},
\vspace{-.1in}
\end{equation}
where $\alpha_{0}$, $\alpha_{1}$, and $\alpha_{2}$ represent the hyperparameters to control the balance amongst the three constituents. In this study, the loss $\mathcal{L}_{\text{refine}}$ is used to directly adjust the parameters of the last layer of the model at each test-time step.  This adjustment preserves the flow structure using LES data, and ensures the consistency with two physics constraints. Due to the auto-regressive nature of the sequential prediction method, the output of the adjusted model can affect the prediction for the following time steps. 

The SR-TR method is different from traditional SR methods in that the low-resolution LES data are only employed in the test-time refinement and are not used in the training. However, it is worth mentioning that  LES data could also be incorporated as additional inputs to 
improve flow reconstruction within the CSTU structure, which will be described later.  



\subsection{Continuous Spatial Representation}

To facilitate flow reconstruction at desired resolutions, the implicit neural representation method \cite{chen2021learning} is integrated into the CSTU structure, as depicted in Fig.~\ref{fig:Structure}.
Before predicting the DNS output $\hat{\textbf{Q}}^d(t+\delta)$ at the target resolution for time $t+\delta$ from the input DNS data, the input $\textbf{Q}^d(t)$ is initially encoded and interpolated to the target resolution of the flow data denoted as $\hat{\textbf{Q}}^\text{c}(t)$. This process is described as:
\vspace{-.05in}
\begin{equation}
\small
\hat{\textbf{Q}}^\text{c}(t) =  {{g}}(t, \textbf{Q}^d(t);\phi),
\vspace{-.05in}
\end{equation}
where ${g}(\cdot)$ contains three components. First, 
it uses CNN layers 
to learn a spatial implicit neural representation. 
The CNN layers convert discrete encoded feature map of each flow data slice to a continuous 2D feature space, where any 2D position can be represented by  its corresponding feature vector $r$. Ideally, the CNN layers need to be trained such that the feature vectors encoded by the CNN layers are uniformly distributed across the 2D space. 

Second, given any query point $x_s$, we select the feature vector $r$ of the grid point $x_z$ that is closest to the queried spatial coordinate $x_s$. This selected vector $r$ is then concatenated with its positional information $x_z$ 
relative to the query point's coordinate $x_s$. 
Finally, this combined input goes through a parameterized interpolation function ${g_s}(\cdot)$ to generate the continuous feature $h_s$ at $x_s$. This process is described as:
\vspace{-.05in}
\begin{equation}
\small
    h_s = {g_s}(r, x_s - x_z).
\vspace{-.05in}
\end{equation}

Next, we use a decoder to convert the continuous feature $h_s$ to the flow values in $\hat{\textbf{Q}}^c$. 
By employing this approach, the CSTU structure can effectively resample the DNS data $\textbf{Q}^d$ to the  DNS data $\hat{\textbf{Q}}^c$ in the desired resolution. 

Additionally, to enhance the capacity of ${g}(\cdot)$ in capturing the feature mapping and improving flow reconstruction across different resolutions, the SR-TR model can be fine-tuned using a small number of DNS samples in  
the target resolution collected from the training period. 
The loss function $\mathcal{L}_\text{finetune}$ between the true HR-DNS data $\textbf{Q}^h$ and the resampled DNS data $\hat{\textbf{Q}}^c$ can be formulated as: 
\vspace{-.05in}
\begin{equation}
\small
    \mathcal{L}_{\text{finetune}} = \text{MSE}(\hat{\textbf{Q}}^c,\textbf{Q}^h).
\vspace{-.05in}
\end{equation}

Once obtaining $\hat{\textbf{Q}}^c$, CSTU utilizes the temporal PRU structure to conduct sequential prediction over time. Different from the original PRU that uses CNNs to approximate the spatial derivatives, CSTU enables estimating spatial derivatives using FDM, as it can interpolate the data at close points to reduce the errors of FDM.  

It is also possible to use the LES data as additional inputs to enhance the intermediate state $\textbf{Q}(t,n)$ in PRU and improve flow reconstruction within the CSTU structure. Same as the PRU architecture shown in Fig.~\ref{fig:PRU}, the initial data point $\textbf{Q}(t) = \hat{\textbf{Q}}^c (t)$ can be replaced by combining DNS and LES data using an augmentation mechanism: $\textbf{Q}(t) = W^c \hat{\textbf{Q}}^c (t) + W^l\textbf{Q}^l (t)$, where $W^c$ and $W^l$ are trainable model parameters. The data $\textbf{Q}^l(t)$ represents the up-sampled LES data with the same resolution as DNS, generated using the ${g}(\cdot)$ method. Following this, the CSTU estimates the initial temporal gradient $\textbf{Q}(t,0)=\textbf{f}(\textbf{Q}(t))$ using the Navier-Stokes equation and calculates the next intermediate state variable $\textbf{Q}(t,1)$ by advancing the flow data $\textbf{Q}(t)$ in the direction of temporal derivatives. With frequent LES data, the intermediate states $\textbf{Q}(t,n)$ are further augmented by incorporating LES data $\textbf{Q}^l(t,n)$, given as $\textbf{Q}(t,n) = W^c \textbf{Q}(t,n) + W^l \textbf{Q}^l(t,n)$. 
For the 4-th order Runga-Kutta method, LES data $\textbf{Q}^l(t,n)$ are selected based on the position of intermediate temporal derivatives computed in the Runga-Kutta method, as $\textbf{Q}^l(t,1)=\textbf{Q}^l(t+\delta/2)$, $\textbf{Q}^l(t,2)=\textbf{Q}^l(t+\delta/2)$, and $\textbf{Q}^l(t,3)=\textbf{Q}^l(t+\delta)$. 
Then CSTU follows a similar process with PRU by moving $\textbf{Q}(t)$ along the estimated gradient $\textbf{Q}_{t,n}$ to compute the subsequent intermediate state $\textbf{Q}(t,n+1)$. 

\section{Experiment}
\subsection{Experimental Settings}

\subsubsection{Dataset} To evaluate the performance of the proposed methodology, the data sets pertaining to two turbulent flows are considered: a forced isotropic turbulent flow (FIT)~\cite{FITdata} and the Taylor-Green vortex (TGV)~\cite{brachet1984taylor} flow. In both scenarios, the mean velocity is zero, $\overline{\bf Q}(t)=0$, and the Reynolds number is sufficiently high to induce turbulent characteristics in the flow.

The FIT dataset comprises the original DNS records of forced isotropic turbulence, representing an incompressible flow. 
The flow is subjected to energy injection at low wave numbers as part of the forcing mechanism. The DNS data consists of $5,024$ time steps, with each step separated by a time interval of $0.002$s, encompassing both velocity and pressure fields. For this study, the original DNS data is generated to three different grids: $128 \times 64 \times 64$, $128 \times 128 \times 128$, and $128 \times 256 \times 256$. Simultaneously, the LES data is generated on grids of size $128 \times 32 \times 32$. Both DNS and LES data are collected along the $128$ equally spaced grid points along the $z$ axis.

The Taylor-Green vortex (TGV) represents another incompressible flow. The evolution of the TGV involves the elongation of vorticity, resulting in the generation of small-scale, dissipating eddies. A box flow scenario is examined within a cubic periodic domain spanning $[-\pi,\pi]$ in all three directions. 
The DNS and LES resolutions are  $ 128 \times 128 \times 65  $ and $ 32 \times 32 \times 65$, respectively. Both of them are produced along the $65$ equally-spaced grid points along the $z$ axis. More details are described in the appendix.

\begin{figure*} [!h]
\vspace{-.2in}
\centering
\subfigure[{$u$ Channel.}]{ \label{fig:a}{}
\includegraphics[width=0.26\linewidth]{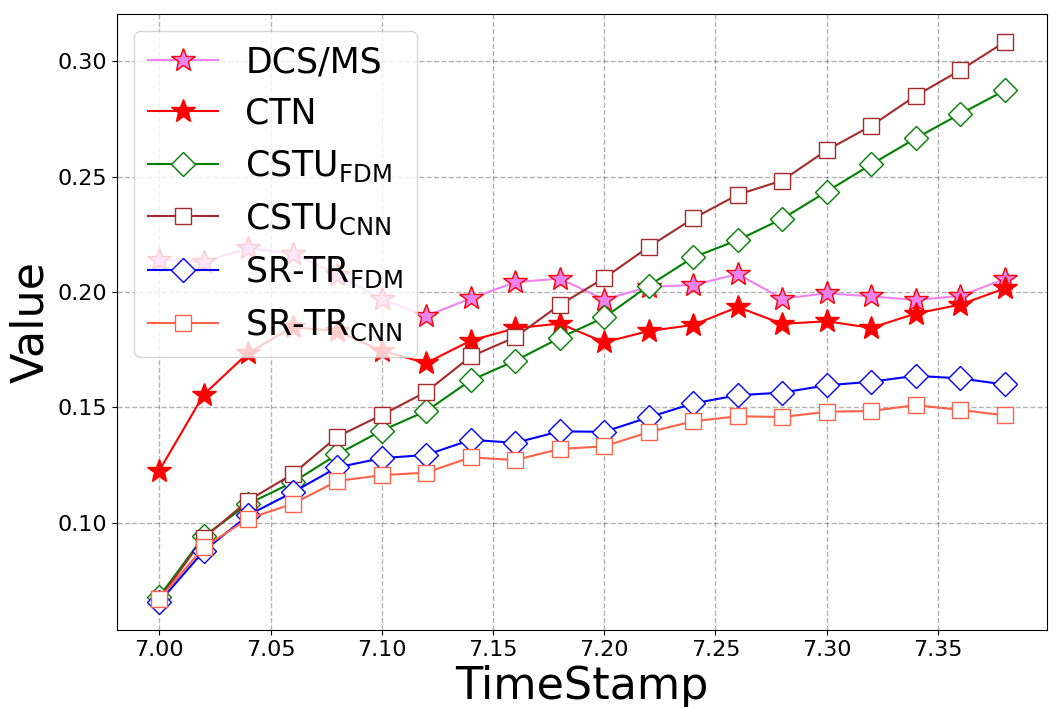}
}
\subfigure[{$v$ Channel.}]{ \label{fig:b}{}
\includegraphics[width=0.26\linewidth]{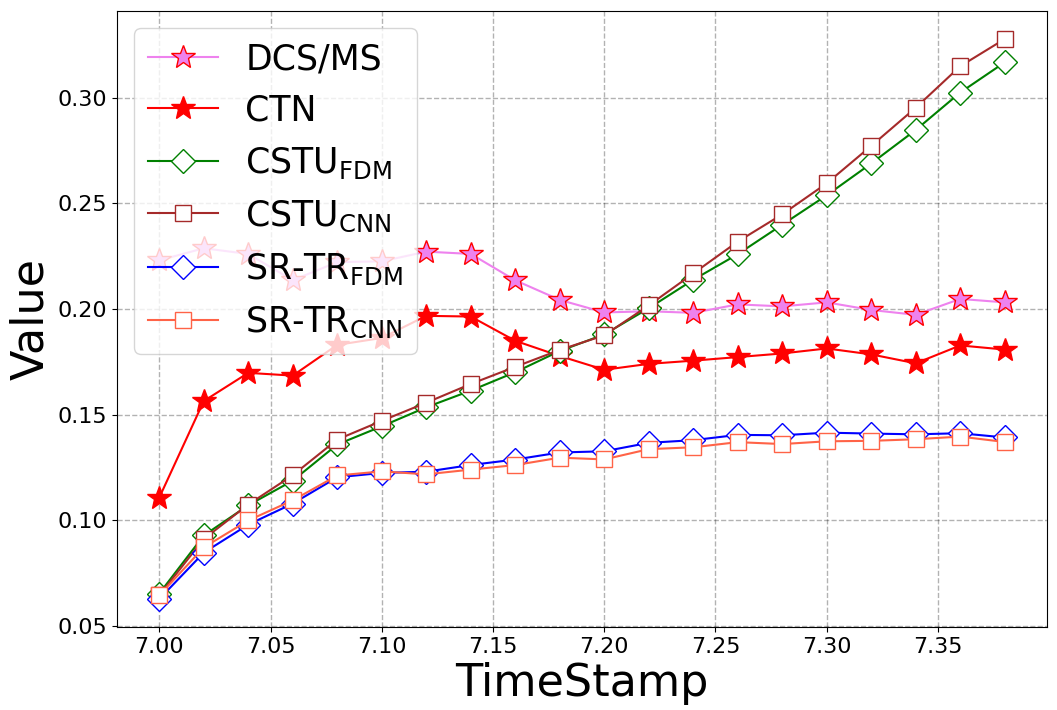}
}
\subfigure[{$w$ Channel.}]{ \label{fig:b}{}
\includegraphics[width=0.26\linewidth]{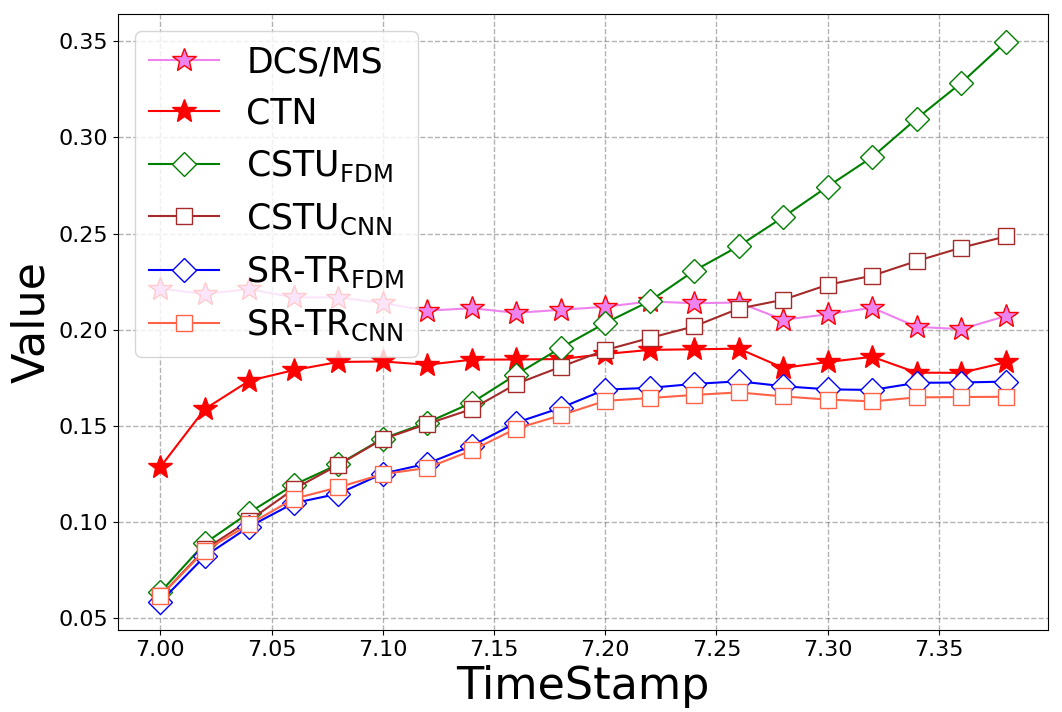}
}
\vspace{-.2in}
\caption{Change of dissipation difference  by different models from 1st (5.6s) to 20th (6s) time step in FIT dataset.}
\label{fig:tf_plot2}
\vspace{-.2in}
\end{figure*}

\begin{figure*} [!h]
\centering
\subfigure[LES Upscaling.]{ \label{fig:a}
\includegraphics[width=0.14\linewidth]{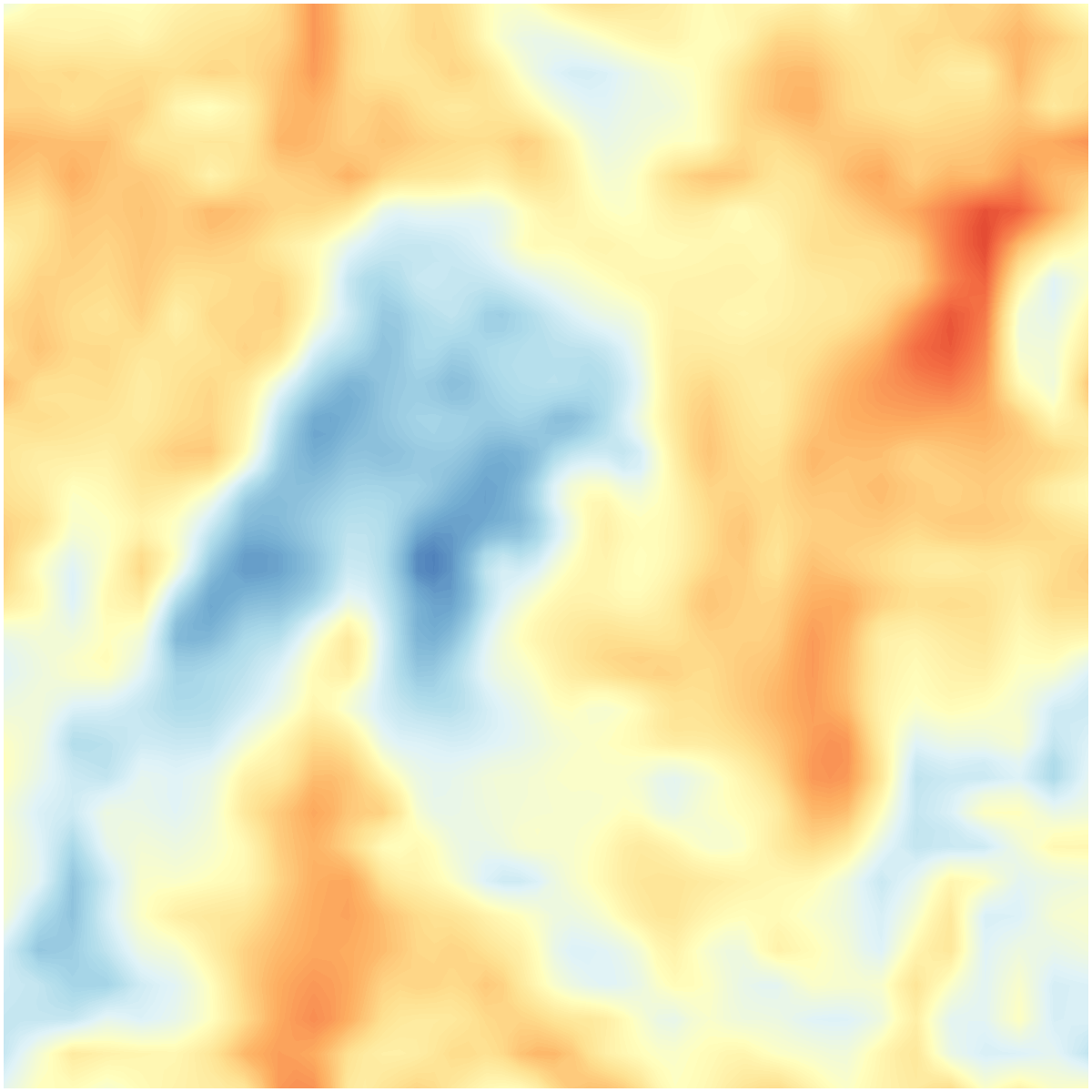}
}
\subfigure[DCS/MS.]{ \label{fig:b}
\includegraphics[width=0.14\linewidth]{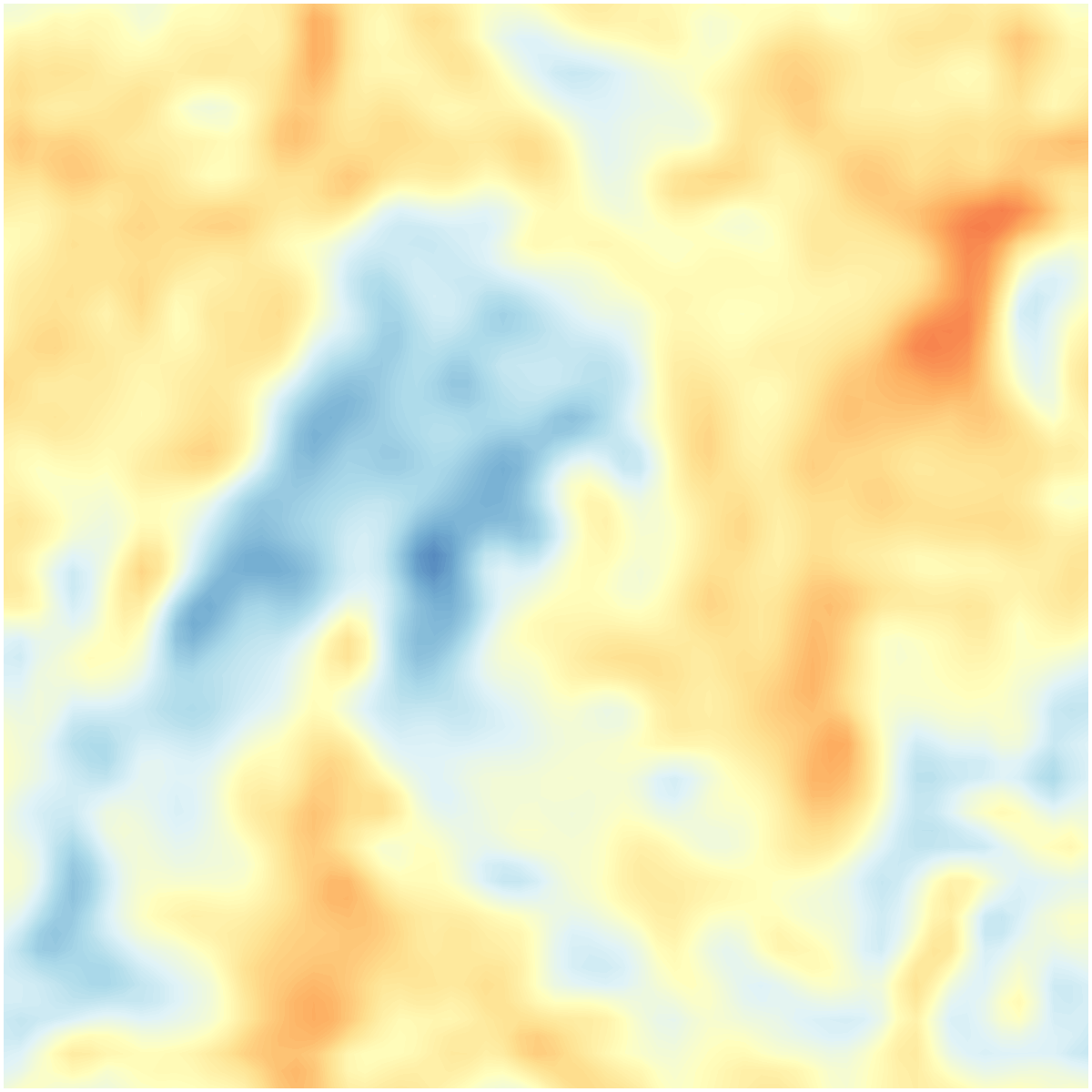}
}
\subfigure[CTN.]{ \label{fig:c}
\includegraphics[width=0.14\linewidth]{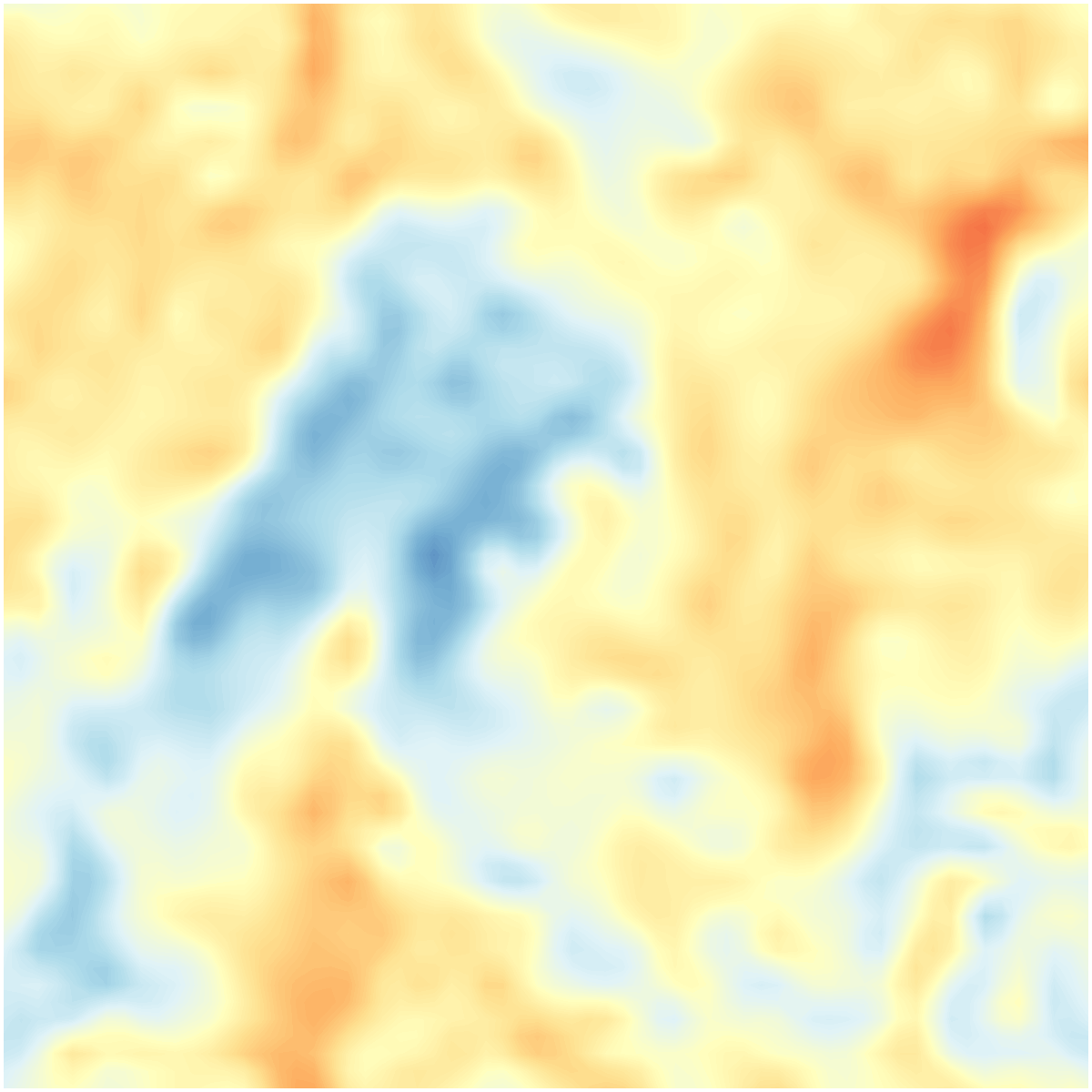}
}
\subfigure[$\text{SR-TR}_\text{FDM}$.]{ \label{fig:d}
\includegraphics[width=0.14\linewidth]{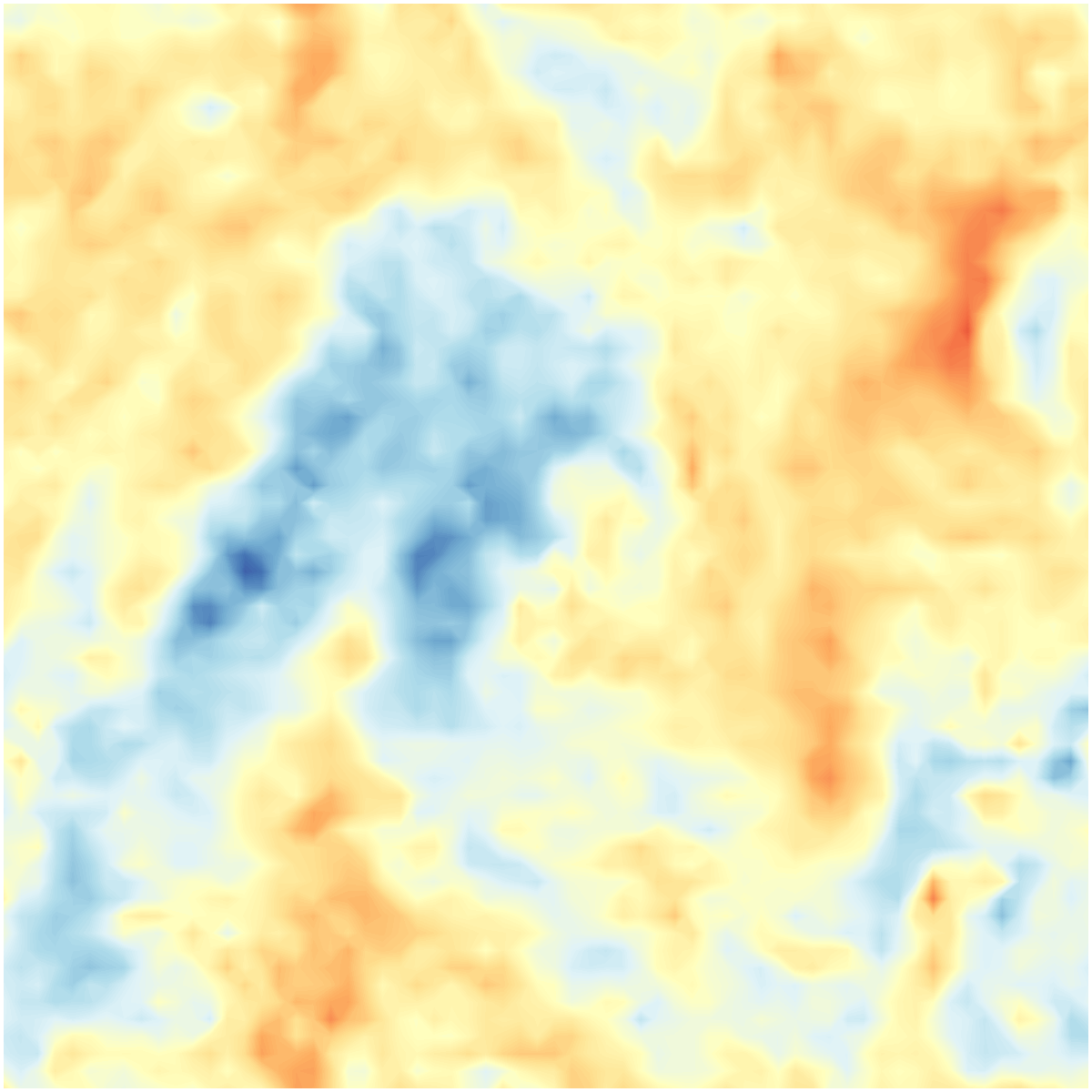}
}
\subfigure[$\text{SR-TR}_\text{CNN}$.]{ \label{fig:e}
\includegraphics[width=0.14\linewidth]{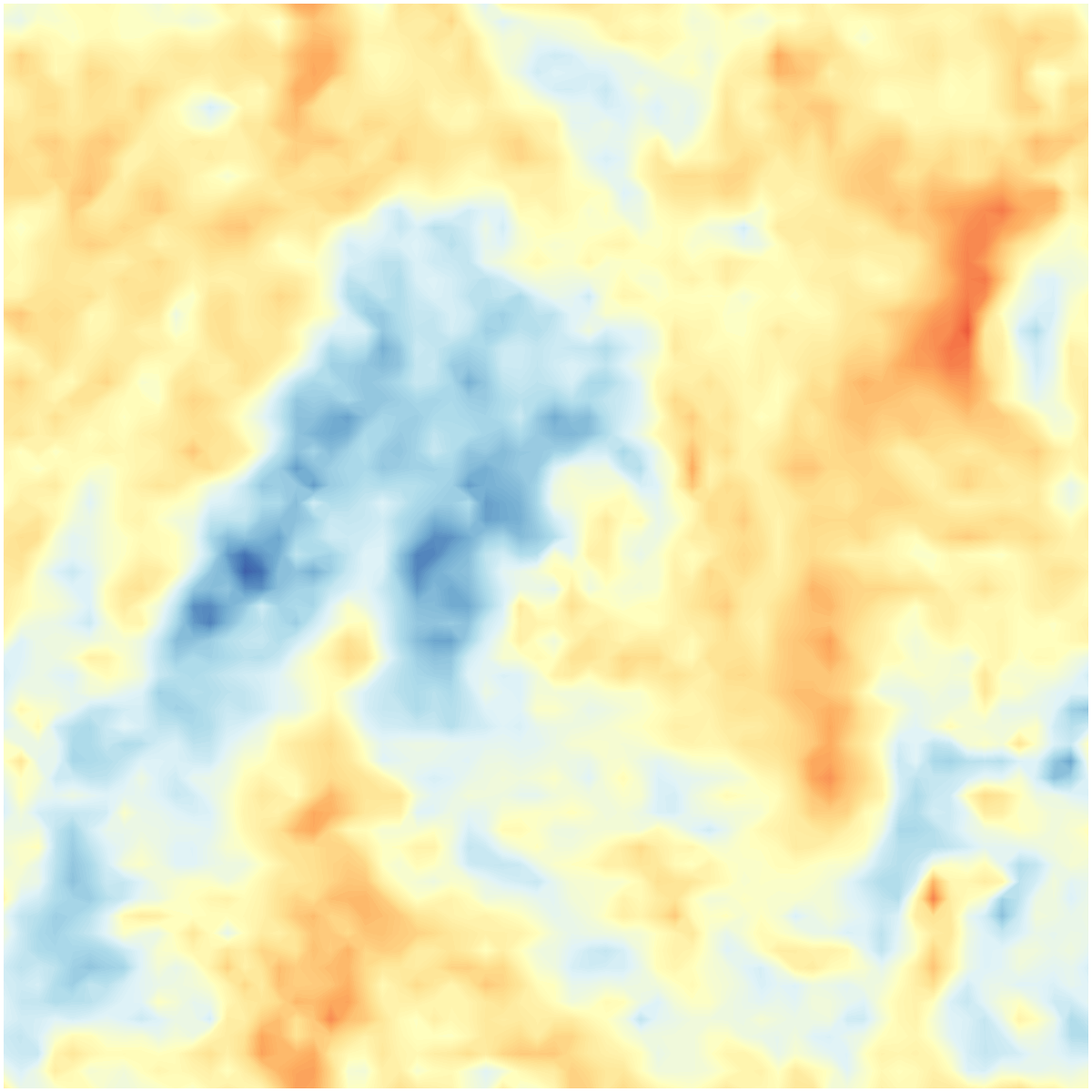}
}
\subfigure[Target DNS.]{ \label{fig:f}
\includegraphics[width=0.14\linewidth]{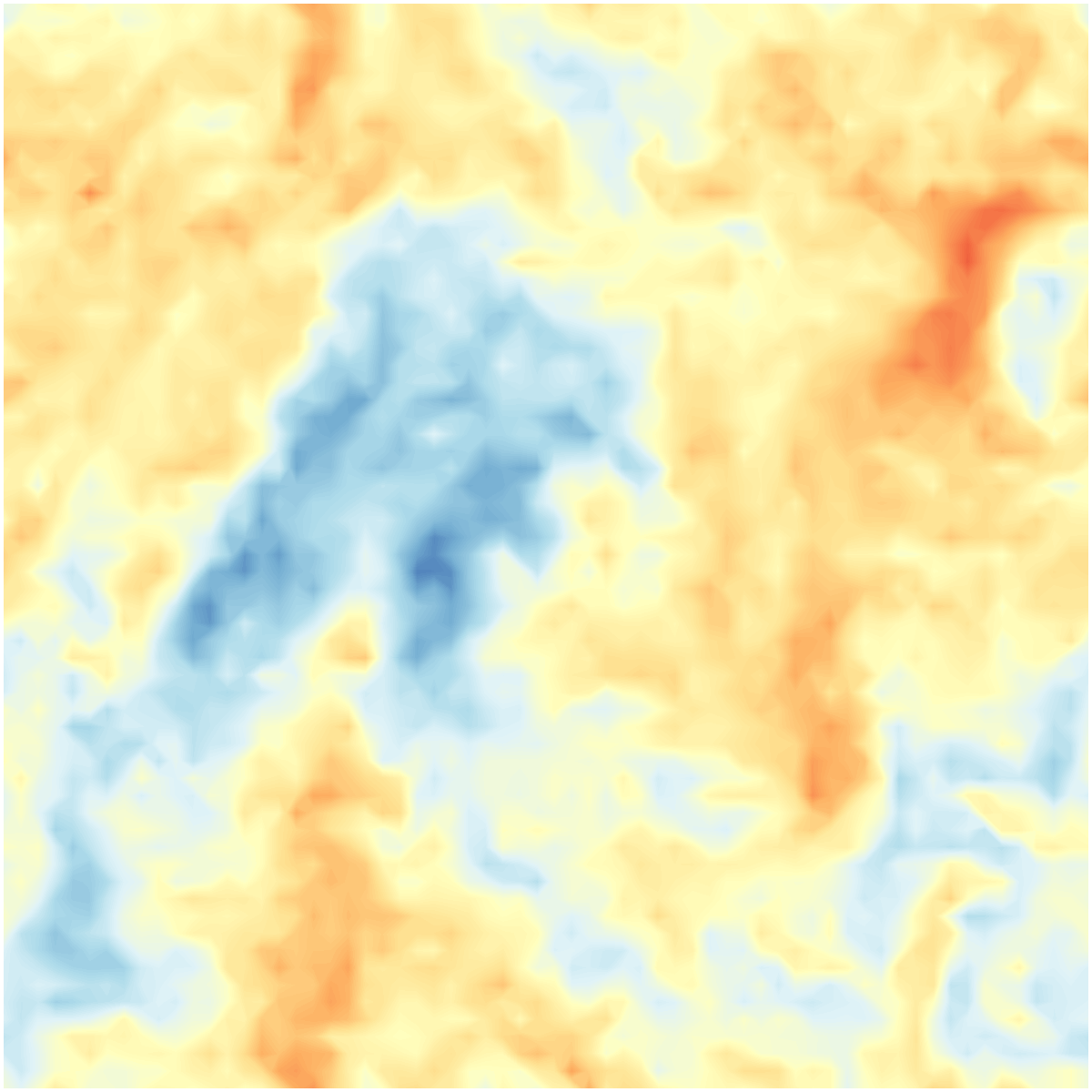}
}
\vspace{-.15in}
\caption{Reconstructed $w$ channel by each method on a sample testing slice along the $z$ dimension in the FIT dataset. The reconstruction results are shown at 20th (6s) in (a)-(f).}
\label{fig:tf_plot3}
\vspace{-.2in}
\end{figure*}

\begin{figure*} [!h]
\centering
\subfigure[LES Upscaling.]{ \label{fig:a}
\includegraphics[width=0.14\linewidth]{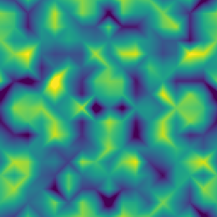}
}
\subfigure[DCS/MS.]{ \label{fig:b}
\includegraphics[width=0.14\linewidth]{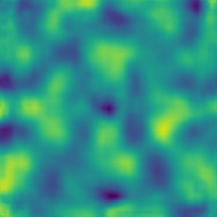}
}
\subfigure[CTN.]{ \label{fig:c}
\includegraphics[width=0.14\linewidth]{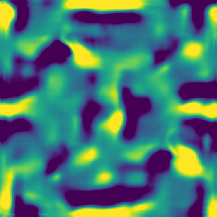}
}
\subfigure[$\text{SR-TR}_\text{FDM}$.]{ \label{fig:d}
\includegraphics[width=0.14\linewidth]{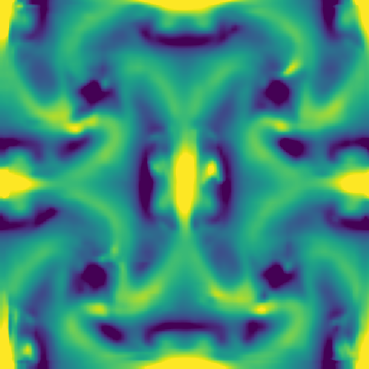}
}
\subfigure[$\text{SR-TR}_\text{CNN}$.]{ \label{fig:e}
\includegraphics[width=0.14\linewidth]{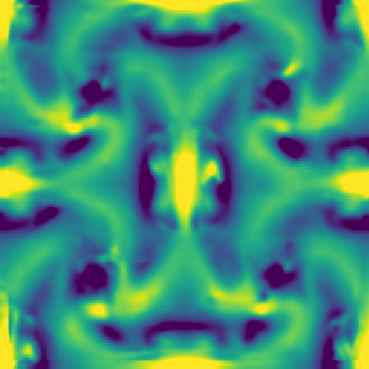}
}
\subfigure[Target DNS.]{ \label{fig:f}
\includegraphics[width=0.14\linewidth]{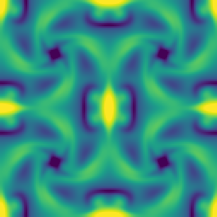}
}
\vspace{-.15in}
\caption{Reconstructed $w$ channel by each method on a sample testing slice along the $z$ dimension in the TGV dataset. The reconstruction results are shown at 15th (110s) in (a)-(f).}
\label{fig:tf_plot4}
\vspace{-.2in}
\end{figure*}

\subsubsection{SR-TR method and baselines} 
The performance of the SR-TR method is evaluated and compared with several existing methods for super-resolution (SR) and turbulent flow downscaling. Specifically, we implement two SR-TR-based methods:  $\text{ST-TR}_\text{FDM}$ and $\text{ST-TR}_\text{CNN}$, which use FDM and CNN to approximate spatial gradients. Additionally, four popular SR methods SRCNN~\cite{dong2014learning}, RCAN~\cite{zhang2018image}, HDRN~\cite{Duong2021}, and SRGAN~\cite{ledig2017photo}, two well-known dynamic fluid downscaling methods: DCS/MS~\cite{fukami2019super} and FSR~\cite{yang2023super}, and the Fourier neural operator (FNO)~\cite{li2020fourier}, are used as baselines.

To better assess the effectiveness of the model components, three additional baseline models are introduced: the Convolutional Transition Network (CTN), $\text{CSTU}_\text{FDM}$, and $\text{CSTU}_\text{CNN}$~\cite{bao2022physics}. The CTN is a combination of SRCNN and  LSTM~\cite{LSTM}. $\text{CSTU}_\text{FDM}$ and $\text{CSTU}_\text{CNN}$ are similar to  $\text{ST-TR}_\text{FDM}$ and $\text{ST-TR}_\text{CNN}$ but do not use degradation-based refinement. The goal of comparing CTN with CSTU-based methods is to highlight the advantages of CSTU in spatio-temporal DNS reconstruction. Comparing CSTU-based methods with SR-TR-based methods shows the benefits of the refinement process. Additionally, we explore the improvement in flow data reconstruction with extra LES input by implementing a variant of the proposed method.

\subsubsection{Experimental designs.} The proposed methods and the baselines are evaluated using both the FIT and TGV datasets. The models are trained using the FIT data over a continuous one-second period with a time interval of $\delta = 0.02s$ and a total of $50$ time steps. Subsequently, the trained models are applied to the following $0.4$ second period (equivalent to $20$ time steps) for performance assessment. For the TGV dataset, the models are trained on a consecutive $40$-second period with a time interval of $\delta = 2s$, and the subsequent $40$ seconds of data (equivalent to $20$ time steps) are used for testing.

Additionally, given that the FIT dataset contains DNS data of different resolutions, it is important to note that all of the methods and baselines were trained using data with a resolution of $64 \times 64$ and were subsequently tested at the same resolution. The higher-resolution DNS data (HR-DNS) are exclusively used for testing the reconstruction of flows at different resolutions and for conducting the ablation study that explores the benefits of utilizing LES data. Additionally, we use the periodic data augmentation~\cite{bao2022physics}  to address boundary conditions.

The assessment of DNS reconstruction performance employs two metrics: the Structural Similarity Index Measure (SSIM)~\cite{wang2004image} and dissipation difference~\cite{enwiki:1127277109}. SSIM measures the similarity between reconstructed and target DNS data in terms of luminance, contrast, and overall structure. Higher SSIM values indicate better reconstruction. Dissipation evaluates the model's gradient capturing ability, considering dissipation for each velocity vector component ($u$, $v$, and $w$). The dissipation operator is defined by $\chi (Q) \equiv \nabla Q \cdot \nabla Q= \left(\frac{\partial Q}{\partial x}\right)^2 + \left(\frac{\partial Q}{\partial y}\right)^2 + \left(\frac{\partial Q}{\partial z}\right)^2$. 
The dissipation is used to measure the difference in flow gradient between the true DNS and generated data. This is  represented by $|\chi({Q}^d) - \chi(\hat{{Q}}^d)|$, and the smaller difference 
indicates better performance.

\subsection{Reconstruction Performance}
\subsubsection{Quantitative results} Table~\ref{fig:table1} and Table~\ref{fig:table2} provide a summary of the average performance across the initial $10$ time steps during the testing phase, tested on both the FIT and TGV datasets. When compared to the baselines, the SR-TR-based methods generally present superior performance in both assessments, exhibiting the highest SSIM value and the lowest dissipation difference.  Several highlights also emerge: (1)~When contrasting SR-TR-based methods with SR baselines, DCS/MS, FSR, and FNO models, it becomes evident that these baseline methods struggle to accurately recover the overall flow, resulting in diminished performance concerning SSIM and dissipation difference. (2)~A comparison between CSTU-based methods and SR-TR-based methods shows the substantial enhancements derived from the incorporation of the degradation-based refinement. 

\begin{table}[!t]
\vspace{-.2in}
\small
\newcommand{\tabincell}[2]{\begin{tabular}{@{}#1@{}}#2\end{tabular}}
\centering
\caption{Reconstruction performance (measured by SSIM, and Dissipation difference) on $(u,v,w)$ channels by different methods in the FIT dataset. The performance is measured by the average results of the first 10 time steps.}
\begin{tabular}{|p{1.4cm}|cc|}
\hline
\textbf{Method} & SSIM $\uparrow$ & Dissipation dif $\downarrow$  \\ \hline 
SRCNN&(0.875, 0.868, 0.871)&(0.235, 0.232, 0.230)\\ 
RCAN& (0.881, 0.871, 0.874)&(0.224, 0.225, 0.225)\\ 
HDRN&(0.887, 0.875, 0.875) &(0.217, 0.223, 0.223)\\
FSR&(0.887, 0.877, 0.875) &(0.218, 0.221, 0.223)\\
DCS/MS&(0.888, 0.878, 0.880) &(0.216, 0.220, 0.214)\\
SRGAN& (0.891, 0.881, 0.215) &(0.215, 0.217, 0.215)\\  
FNO&(0.912, 0.915, 0.911) &(0.153, 0.151, 0.150)\\
CTN& (0.901, 0.891, 0.903) &(0.161, 0.173, 0.174)\\  
\hline
$\text{CSTU}_\text{FDM}$&(0.936, 0.941, 0.944)&(0.131, 0.133, 0.132)\\ 
$\text{CSTU}_\text{CNN}$&(0.935, 0.943, 0.945) &(0.134, 0.134, 0.129)\\
$\text{SR-TR}_\text{FDM}$&(0.949, 0.953, 0.951) &(0.111, 0.111, 0.116)\\
$\text{SR-TR}_\text{CNN}$&(0.950, 0.953, 0.952) &(0.116, 0.112, 0.115)\\
\hline
\end{tabular}
\label{fig:table1}
\end{table}

\begin{figure} [!t] 
\centering
\includegraphics[width=0.6\columnwidth]{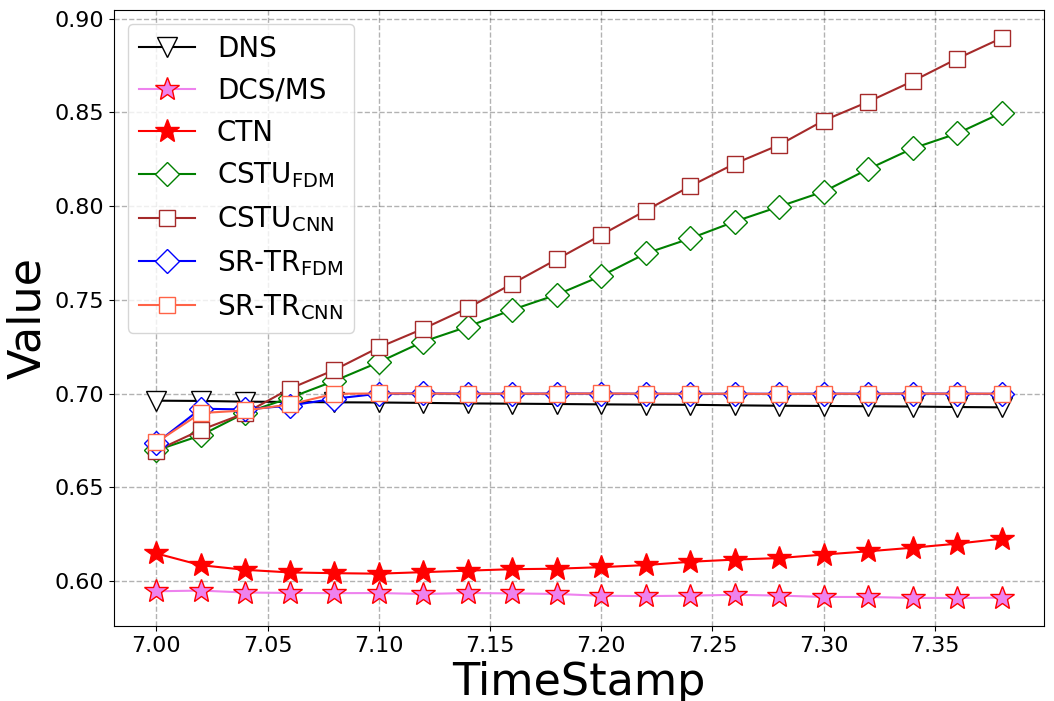}
\vspace{-0.1in}
\caption{Change of kinetic energy produced by the reference DNS and different models in the FIT datasets.}
\label{fig:kinetic}
\end{figure}

\begin{table}[!t]
\vspace{-.2in}
\small
\newcommand{\tabincell}[2]{\begin{tabular}{@{}#1@{}}#2\end{tabular}}
\centering
\caption{Reconstruction performance (measured by SSIM, and Dissipation difference) on $(u,v,w)$ channels by different methods in the TGV dataset. The performance is measured by the average results of the first 10 time steps.}
\begin{tabular}{|p{1.4cm}|cc|}
\hline
\textbf{Method} &  SSIM $\uparrow$ & Dissipation dif$\times$10 $\downarrow$ \\ \hline 
SRCNN&(0.602, 0.603, 0.626) &(0.083, 0.087, 0.079)\\ 
RCAN& (0.627, 0.622, 0.631)&(0.073, 0.074, 0.071)\\ 
HDRN&(0.638, 0.638, 0.641) &(0.072, 0.072, 0.068)\\
FSR&(0.646, 0.648, 0.649) &(0.070, 0.073, 0.066)\\
DSC/MS&(0.647, 0.649, 0.649) &(0.070, 0.071, 0.065)\\
SRGAN& (0.661, 0.658, 0.666) &(0.068, 0.067,0.058)\\
FNO&(0.645, 0.646, 0.648) &(0.072, 0.071, 0.072)\\
CTN& (0.623, 0.624, 0.627) &(0.093, 0.096, 0.087)\\  
\hline
$\text{CSTU}_\text{FDM}$&(0.705, 0.708, 0.701)&(0.049, 0.045, 0.043)\\ 
$\text{CSTU}_\text{CNN}$&(0.708, 0.705, 0.702) &(0.048, 0.046, 0.043)\\
$\text{SR-TR}_\text{FDM}$&(0.918, 0.919, 0.878) &(0.032, 0.032, 0.026)\\
$\text{SR-TR}_\text{CNN}$&(0.913, 0.916, 0.874) &(0.033, 0.034, 0.026)\\
\hline
\end{tabular}
\label{fig:table2}
\end{table}

\subsubsection{Temporal analysis and visualization} 
For the FIT dataset, the performance for reconstructing DNS is measured for each step during a $0.4s$ period ($20$ time steps) in the testing phase. The performance change using dissipation difference is shown in  Fig.~\ref{fig:tf_plot2}. Several observations are highlighted: (1)~With larger time intervals between training and prediction data, the performance becomes worse. In general, SR-TR-based methods exhibit greater stability over long-term prediction, indicating superior performance compared to other methods. (2)~SR-TR-based methods outperform CSTU-based methods, illustrating the effectiveness of degradation-based refinement in mitigating prediction bias over long-term predictions. (3)~Both $\text{CSTU}_\text{FDM}$ and $\text{CSTU}_\text{CNN}$ demonstrate similar performance. A parallel observation arises when comparing the two variants of SR-TR-based methods, demonstrating that either approach for estimating the spatial derivative within the CSTU can obtain similar performance. The same conclusion can be drawn from the temporal analysis of the TGV data, shown in appendix. 

The effectiveness of the SR-TR method is further supported by the visual results shown in Fig.~\ref{fig:tf_plot3}, which illustrates the flow reconstruction performance at the 20th time step after the training phase. The presented slices depict the $w$ component at specific $z$ values. Most of the baseline methods struggle to capture the correct flow transport patterns. In contrast, SR-TR-based methods exhibit significantly improved performance in the later stages. Similar trends are observed in the TGV dataset, as shown in Fig.~\ref{fig:tf_plot4}. 

\subsubsection{Validation via physical metrics.} 
The performance is also evaluated through the long-term prediction of turbulent kinetic energy. Figure~\ref{fig:kinetic} illustrates the energies corresponding to the target DNS, along with the reconstructed flow data from both the baselines and SR-TR-based methods for the FIT dataset. The following observations can be made: (1)~The SR-TR-based methods exhibit superior performance compared to the baseline methods DCS/MS and CTN. (2)~The performance of the CSTU-based methods degrades significantly after the 10th time step. This amplification of accumulated errors at each time step contributes to this outcome. A similar conclusion can be drawn from the analysis of TGV data in appendix. 

\begin{figure} [!t] 
\vspace{-.1in}
\centering
\subfigure[SSIM.]{ \label{fig:b}{}
\includegraphics[width=0.35\columnwidth]{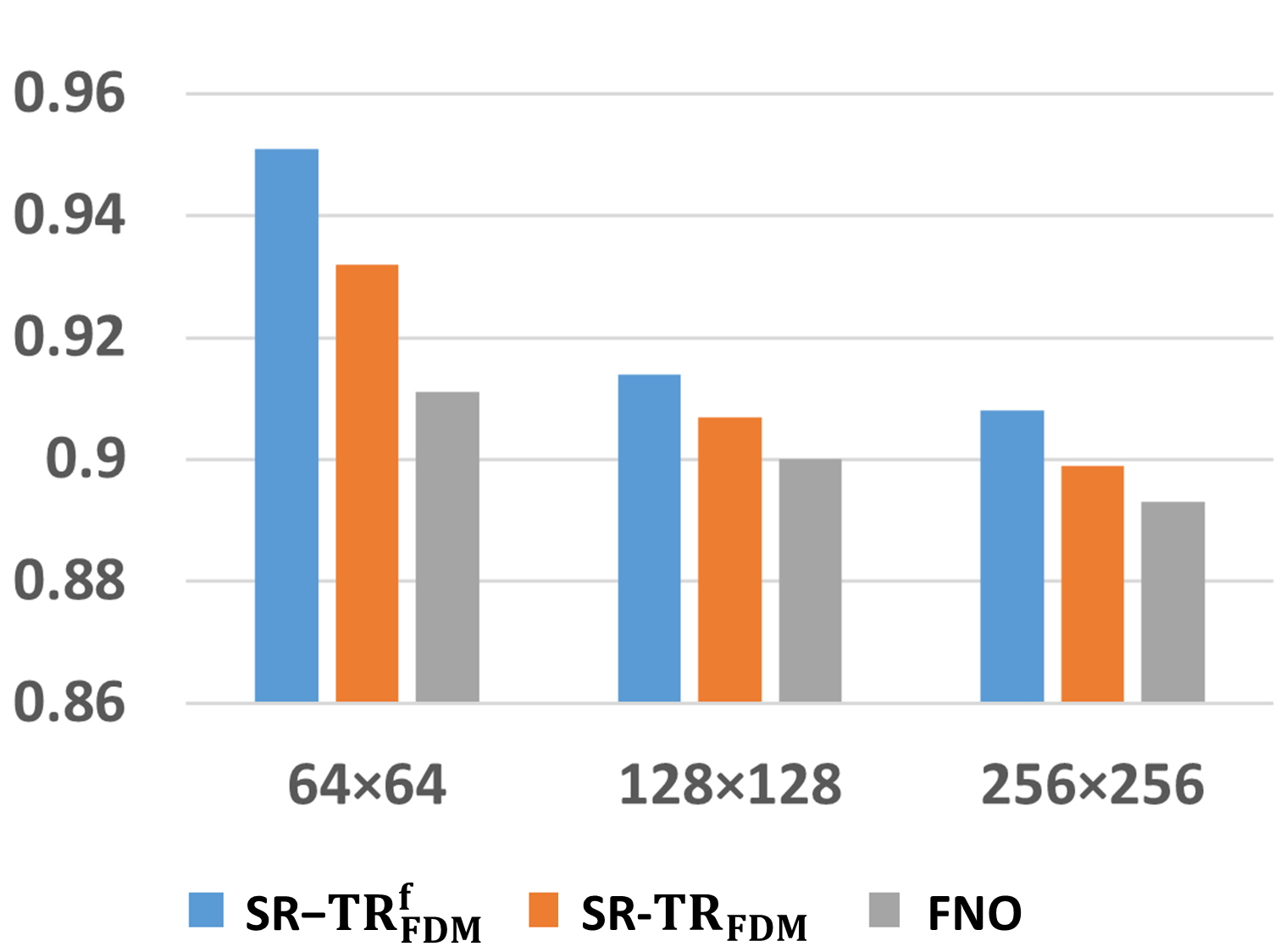}
}\hspace{.1in}
\subfigure[Dissipation diff.]{ \label{fig:b}{}
\includegraphics[width=0.35\columnwidth]{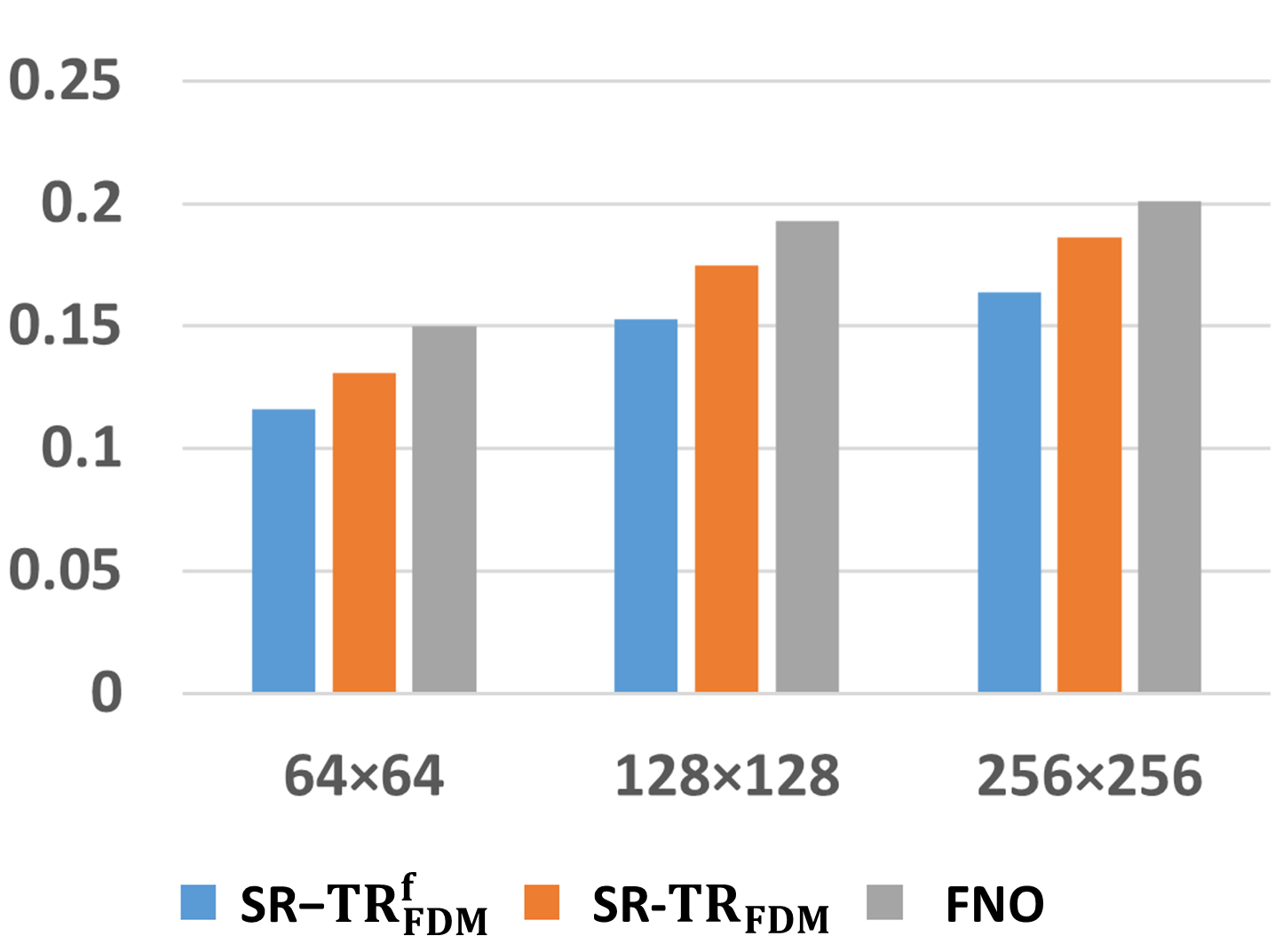}
}\vspace{-.15in}
\caption{The quantitative performance of the models in the $w$ channel is evaluated across different resolutions. Specifically, FNO and $\text{SR-TR}_\text{FDM}$ are purely zero-shot super-resolution methods. $\text{SR-TR}_\text{FDM}^\text{f}$ utilizes  higher resolution DNS data at a few steps for model fine-tuning before prediction. }
\label{fig:resolution}
\end{figure}

\subsubsection{Reconstruction performance in different resolutions.} 
We compare the performance of $\text{SR-TR}_\text{FDM}$, $\text{SR-TR}_\text{FDM}^\text{f}$, and  FNO on the FIT dataset. All models are trained at a resolution of $64 \times 64$, tested at different resolutions: $64 \times 64$, $128 \times 128$, and $256 \times 256$. Notably, FNO and $\text{SR-TR}_\text{FDM}$ are purely zero-shot super-resolution~\cite{shocher2018zero}  methods, without using higher resolution DNS data for model fine-tuning before prediction. While $\text{SR-TR}_\text{FDM}^\text{f}$ utilizes 15 time steps' higher resolution DNS data for model fine-tuning before prediction. Figure~\ref{fig:resolution} displays the performance of three methods, highlighting some observations: (1)~All methods face more difficulty recovering higher-resolution flow data. This is due to the increased complexity of flow details in higher-resolution data. 
(2)~When comparing the performance between $\text{SR-TR}_\text{FDM}$ and $\text{SR-TR}_\text{FDM}^\text{f}$, it is obvious that fine-tuning the model before prediction can lead to significant improvements in flow reconstruction across different higher resolutions. (3)~$\text{SR-TR}_\text{FDM}$ method can achieve ideal reconstruction performance and demonstrate zero-shot super-resolution capabilities based on SSIM and dissipation difference. In contrast, FNO struggles to make accurate predictions at the same resolution and also faces challenges in generalizing to unseen resolutions. These observations highlight the effectiveness of $\text{SR-TR}_\text{FDM}$ in long-term reconstruction and zero-shot super-resolution. 

\begin{figure} [!h] 
\vspace{-.1in}
\centering
\subfigure[SSIM.]{ \label{fig:b}{}
\includegraphics[width=0.35\columnwidth]{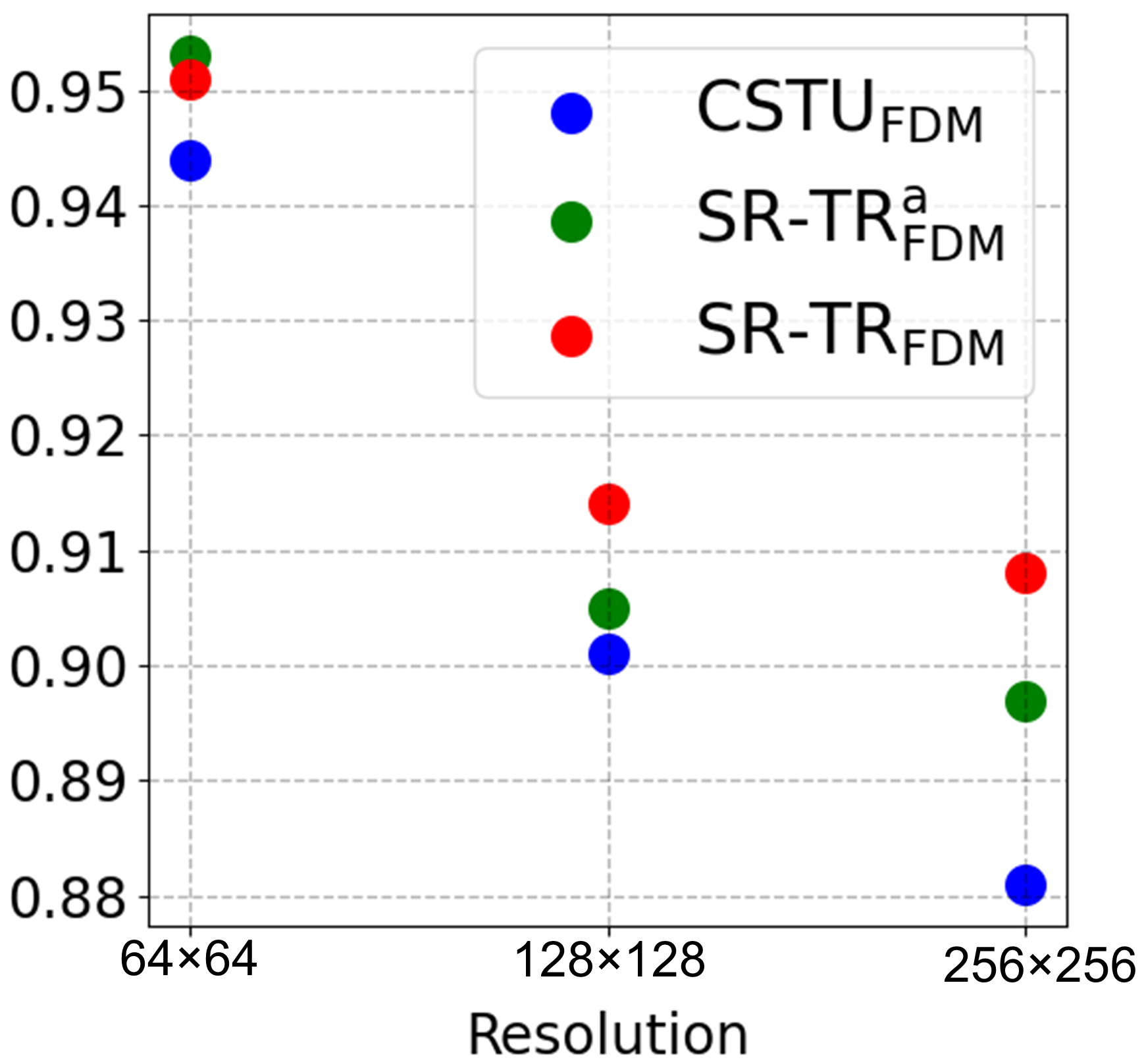}
}\hspace{.1in}
\subfigure[Dissipation diff.]{ \label{fig:b}
\includegraphics[width=0.34\columnwidth]{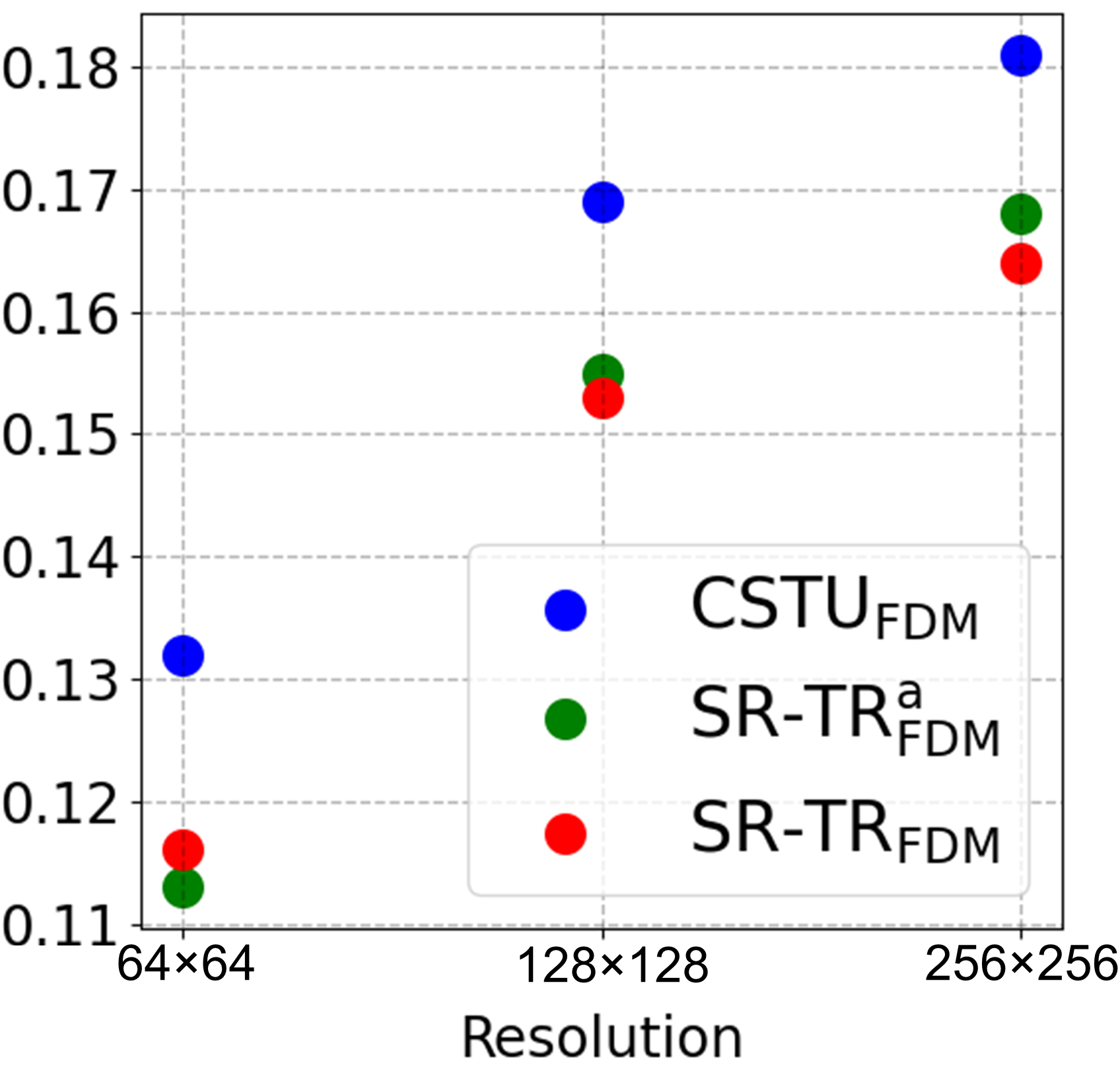}
}\vspace{-.15in}
\caption{The quantitative performance of the models in the $w$ channel is evaluated as the utilization of LES data varies. Specifically, $\text{CSTU}_\text{FDM}$ does not incorporate LES data, $\text{SR-TR}_\text{FDM}$ only leverages LES data in the refinement process, and $\text{SR-TR}_\text{FDM}^\text{a}$ utilizes LES data both as input and in the refinement process.
}
\label{fig:ablation}
\end{figure}

\subsubsection{Ablation study for utilization of LES data.} The objective of this study is to investigate techniques for integrating LES data into the SR-TR approach. The performance of these techniques is shown in Fig.~\ref{fig:ablation}, generally demonstrating their effectiveness in utilizing LES data for reconstructing flow data across different resolutions. Several observations can be highlighted: (1)~When comparing $\text{CSTU}_\text{FDM}$ with $\text{SR-TR}_\text{FDM}$ or $\text{SR-TR}_\text{FDM}^\text{a}$, a significant improvement in reconstruction performance, as indicated by both SSIM and dissipation difference, can be observed. This demonstrates that introducing LES data can lead to improvements regardless of the specific LES data utilization strategy employed.
(2)~Compared with $\text{SR-TR}_\text{FDM}$,  $\text{SR-TR}_\text{FDM}^\text{a}$ method demonstrates slightly better performance in reconstructing flow data at the same resolution. However, its performance is worse than that of $\text{SR-TR}_\text{FDM}$ when aiming to reconstruct higher-resolution flow data. This discrepancy comes from the significant dissimilarity between low-resolution LES data and high-resolution DNS data, which not only hinders the reconstruction process but can also lead to performance degradation, particularly for higher resolutions. Based 
on these observations, we conclude that 
using the low-resolution LES data as input does not bring significant benefit when 
the degradation-based refinement process is adopted.

\section{Conclusion}


A new physics-guided neural network, called "super-resolution through test time refinement" (SR-TR), has been developed to reconstruct high-resolution flow data at various resolutions and time intervals. SR-TR is designed for unsteady, incompressible, Newtonian turbulent flow in spatially homogeneous conditions. The key component, the Continuous Spatial Transition Unit (CSTU), leverages physical insights from the Navier-Stokes equation to capture spatial and temporal flow dynamics. CSTU also enables reconstruction across different resolutions. To enhance the precision of the reconstructed data over time, a degradation-based refinement method is introduced, ensuring the data remains true to physical constraints. The model's performance is tested on two turbulent flow scenarios, using both visual and statistical analysis, showing SR-TR's superior spatio-temporal reconstruction ability. Notably, the constituents of the model, CSTU and the degradation-based refinement, can easily serve as fundamental building blocks for enhancing existing deep learning models.

\bibliographystyle{unsrt}
\bibliography{reference}

\appendix

\newpage

\begin{figure*} [!h]
\centering
\subfigure[{$u$ Channel.}]{ \label{fig:a}{}
\includegraphics[width=0.3\linewidth]{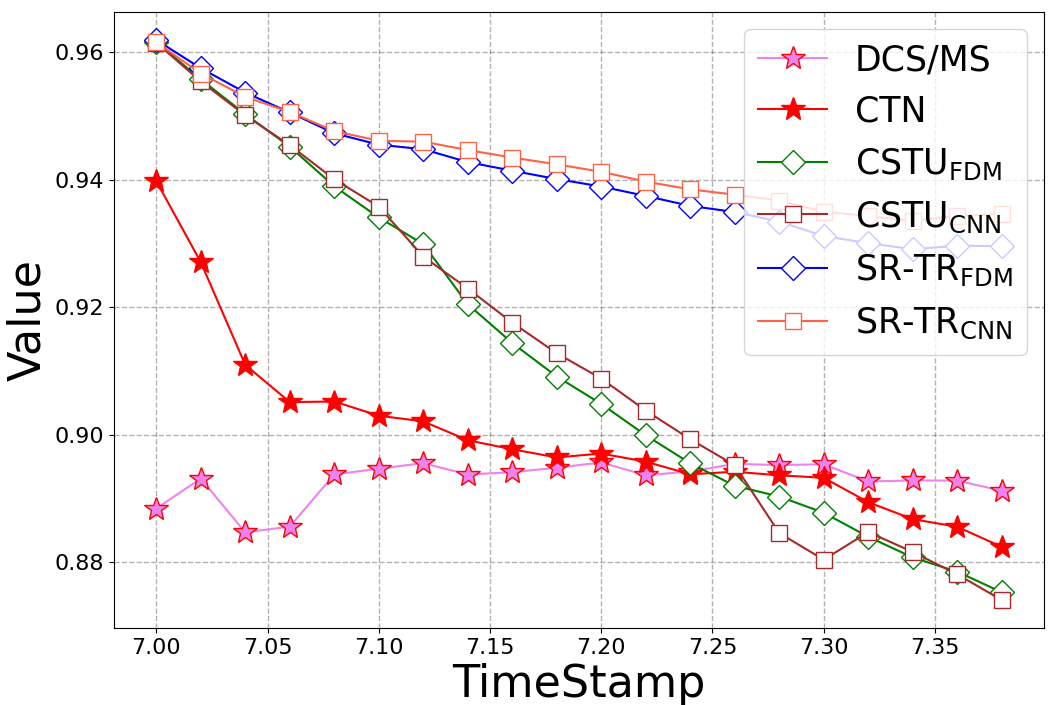}
}
\subfigure[{$v$ Channel.}]{ \label{fig:b}{}
\includegraphics[width=0.3\linewidth]{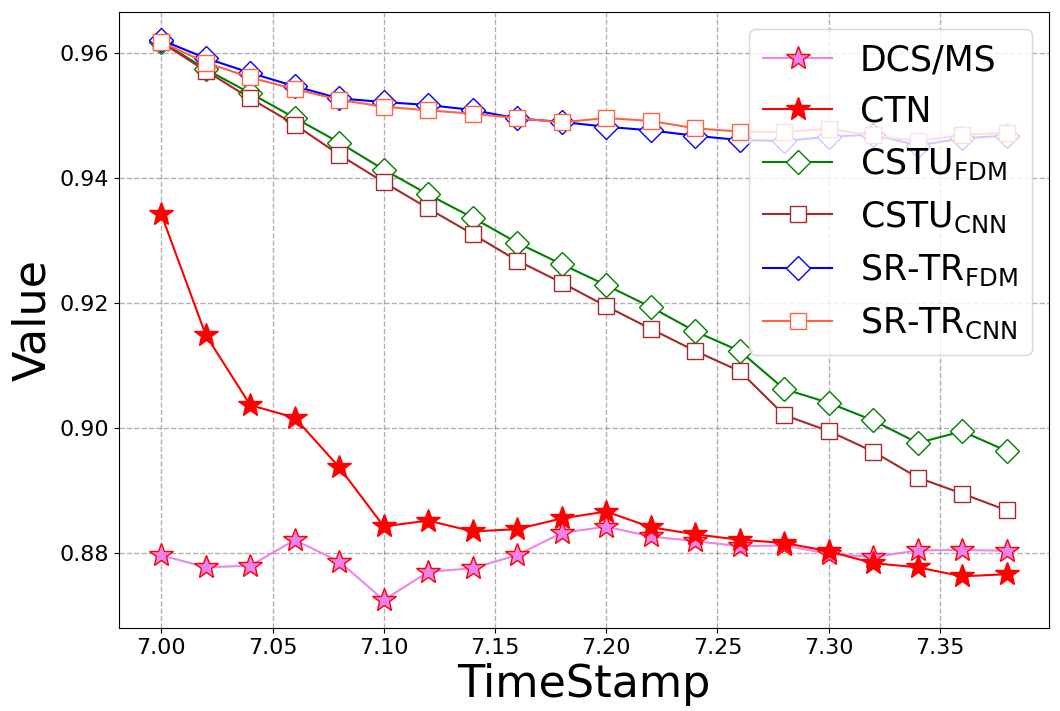}
}
\subfigure[{$w$ Channel.}]{ \label{fig:b}{}
\includegraphics[width=0.3\linewidth]{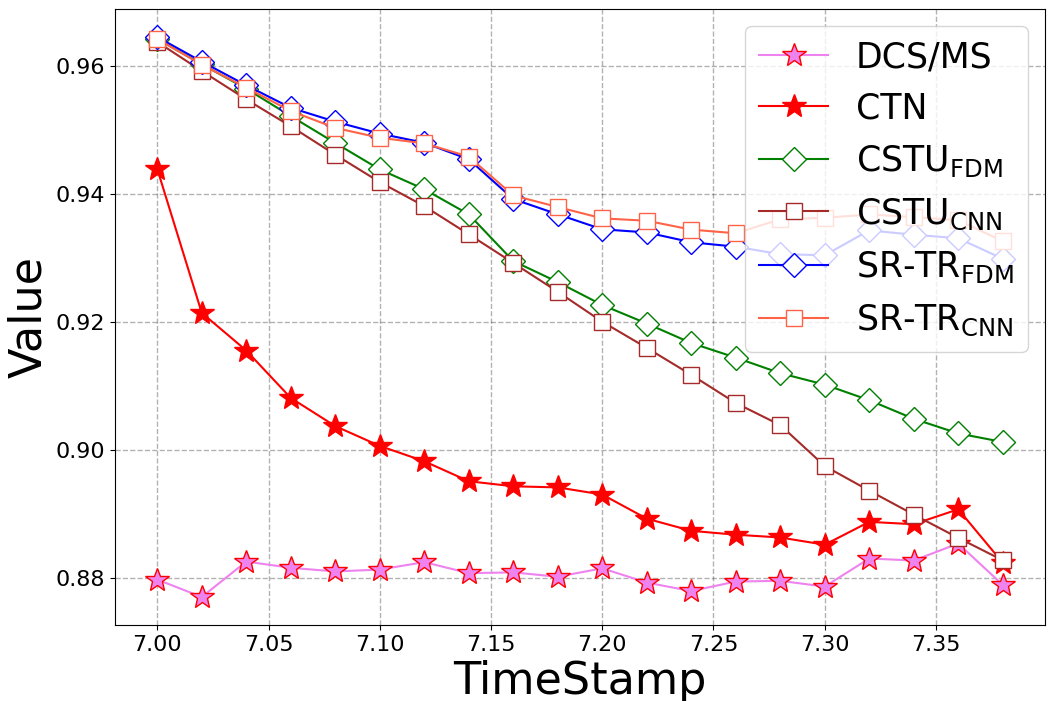}
}
\vspace{-.1in}
\caption{Change of SSIM produced by different models from 1st (5.6s) to 20th (6s) time step in FIT dataset.}
\label{fig:tf_plot1}
\vspace{-.15in}
\end{figure*}

\begin{figure*} [!h]
\centering
\subfigure[{$u$ Channel.}]{ \label{fig:a}{}
\includegraphics[width=0.3\linewidth]{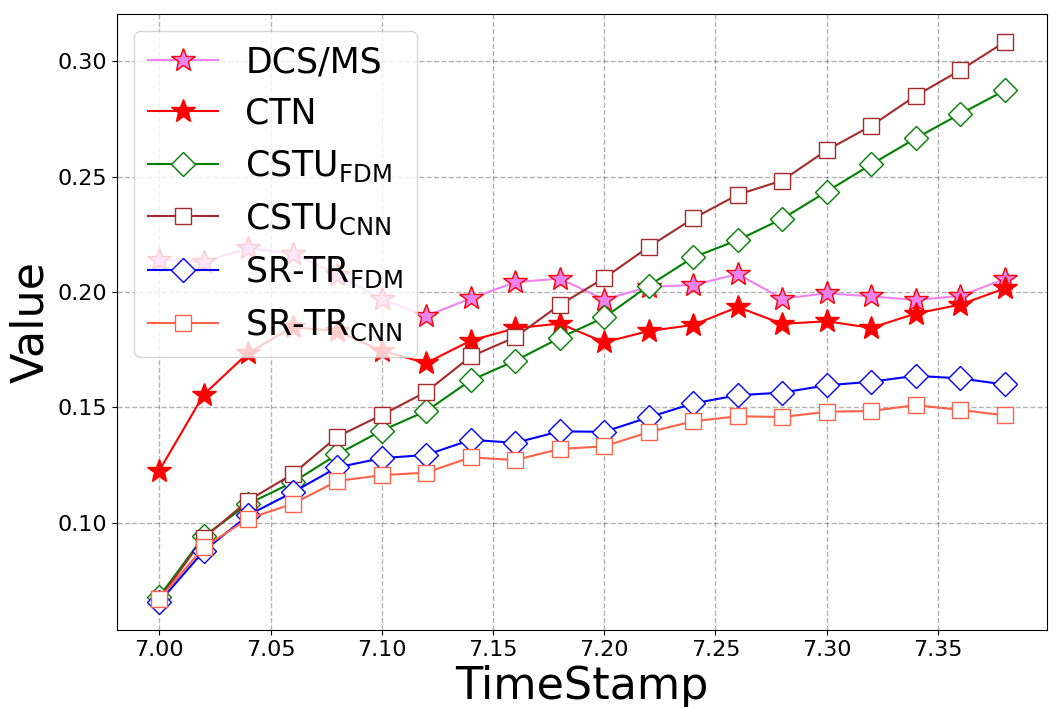}
}
\subfigure[{$v$ Channel.}]{ \label{fig:b}{}
\includegraphics[width=0.3\linewidth]{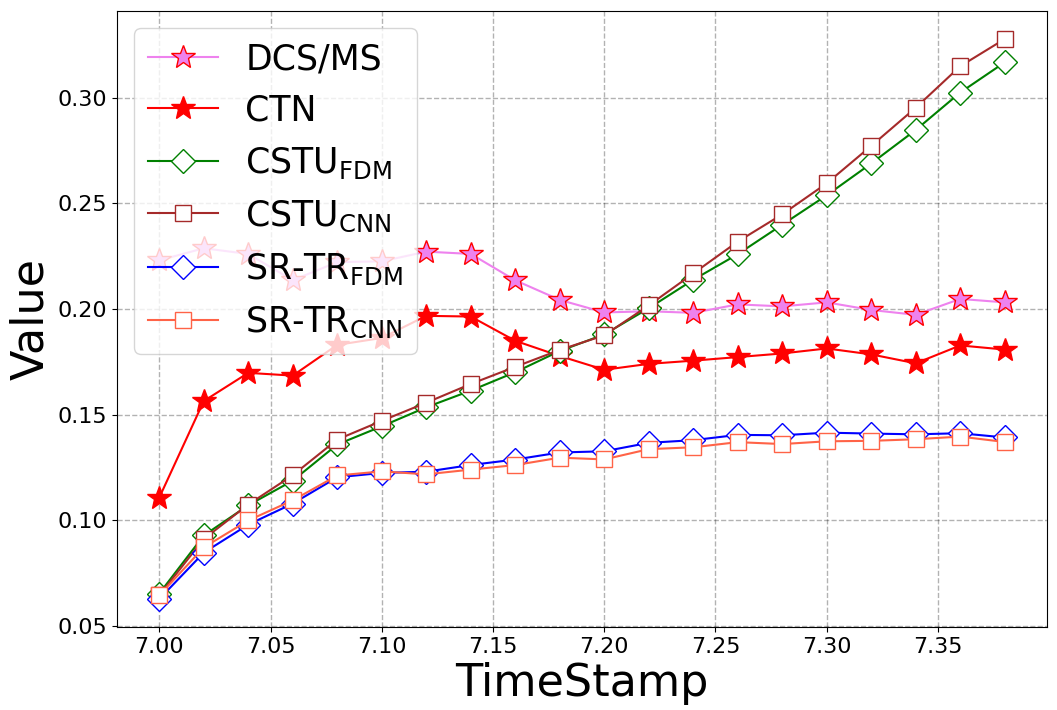}
}
\subfigure[{$w$ Channel.}]{ \label{fig:b}{}
\includegraphics[width=0.3\linewidth]{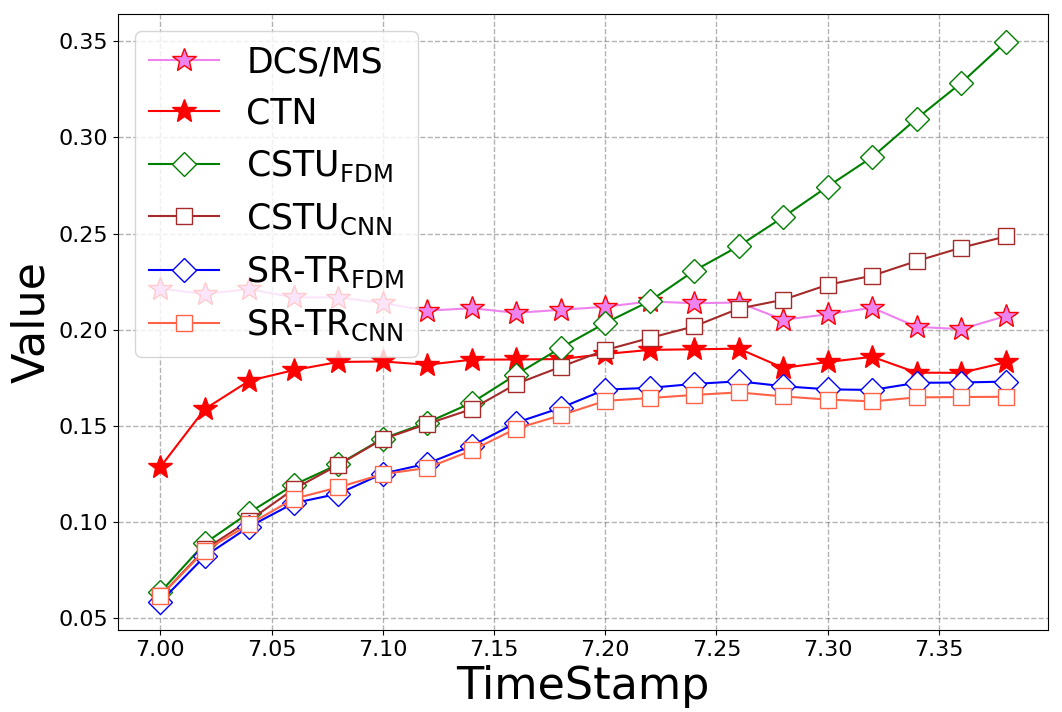}
}
\vspace{-.1in}
\caption{Change of dissipation difference  by different models from 1st (5.6s) to 20th (6s) time step in FIT dataset.}
\label{fig:tf_plot2}
\vspace{-.15in}
\end{figure*}

\begin{figure*} [!h]
\centering
\subfigure[{$u$ Channel.}]{ \label{fig:a}{}
\includegraphics[width=0.3\linewidth]{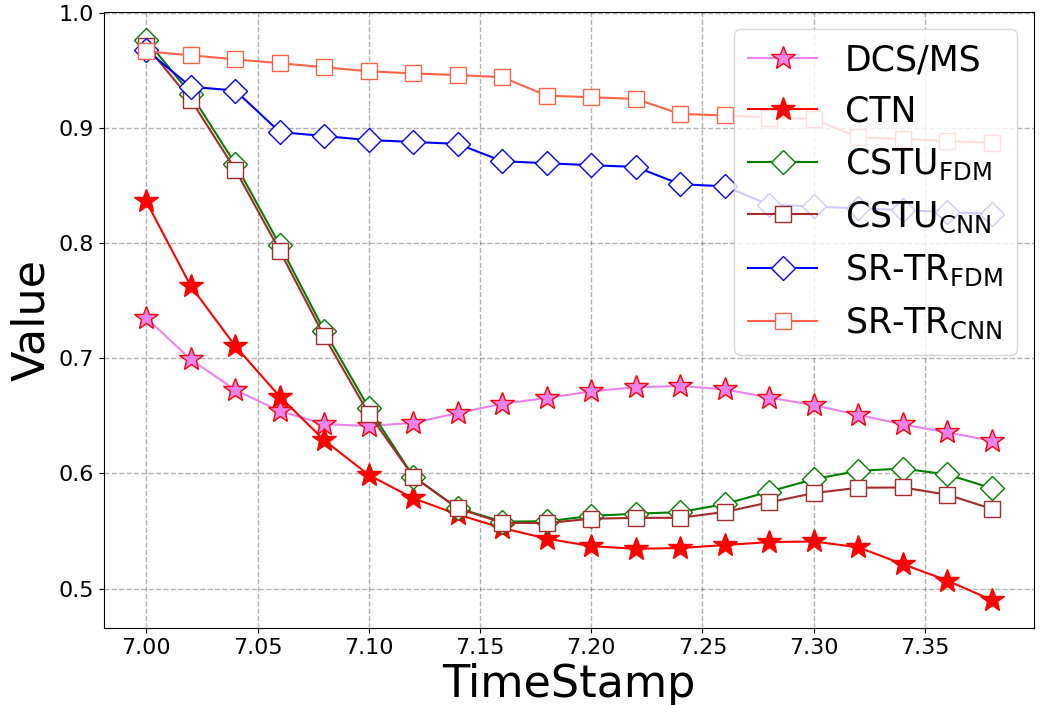}
}
\subfigure[{$v$ Channel.}]{ \label{fig:b}{}
\includegraphics[width=0.3\linewidth]{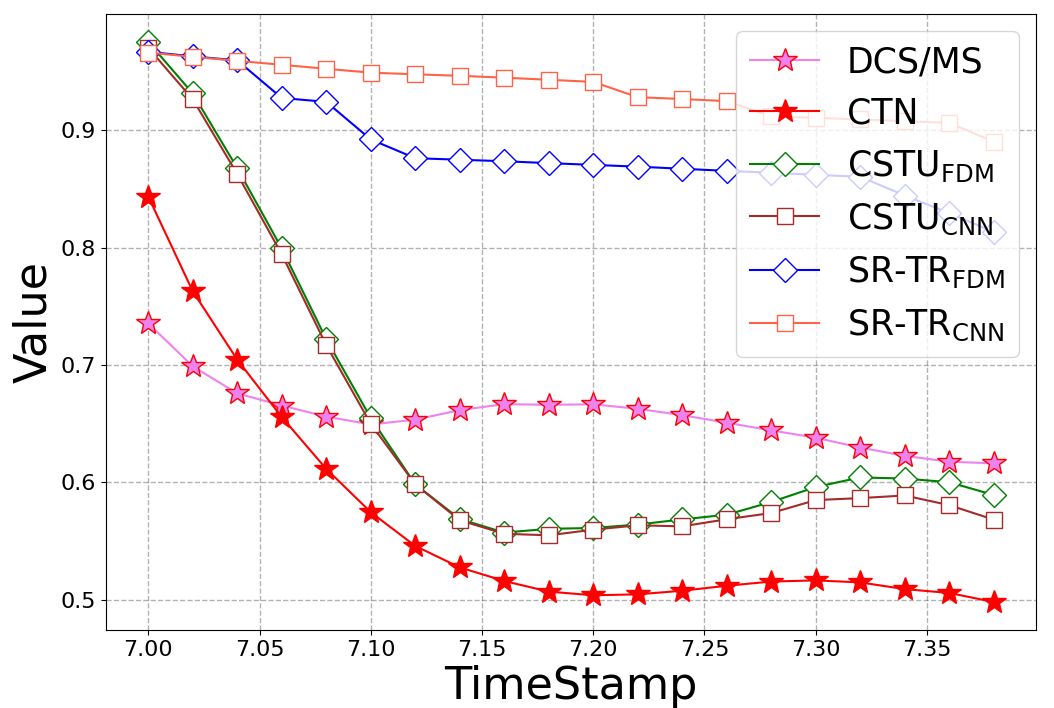}
}
\subfigure[{$w$ Channel.}]{ \label{fig:b}{}
\includegraphics[width=0.3\linewidth]{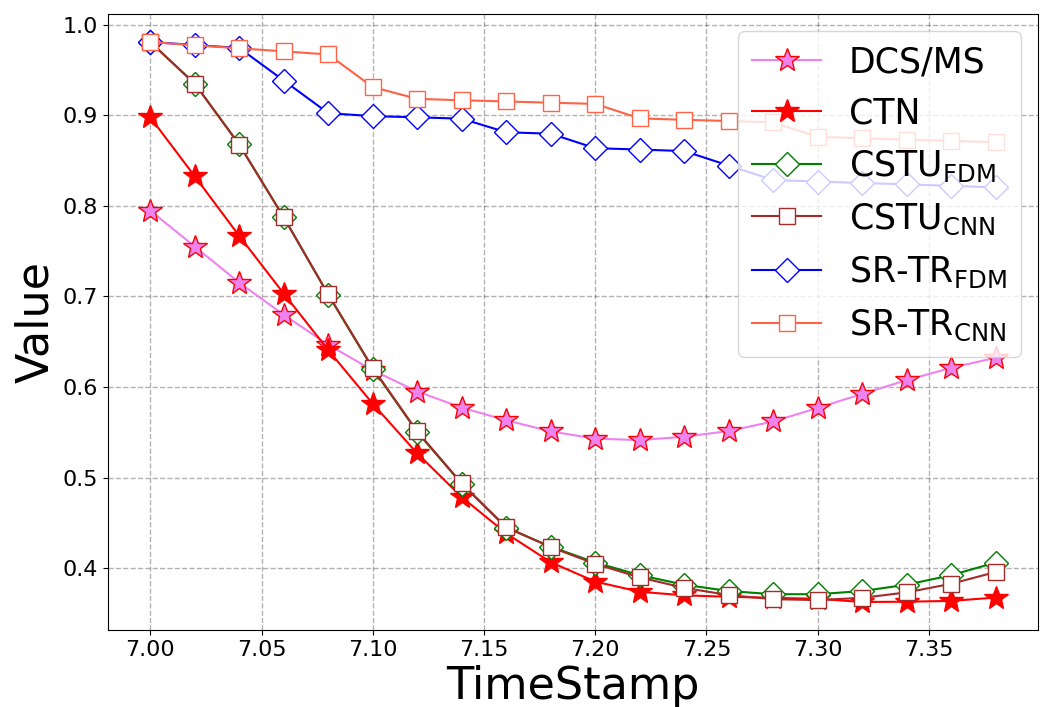}
}
\vspace{-.1in}
\caption{Change of SSIM produced by different models from 1st (80s) to 20th (120s) time step in TGV dataset.}
\label{fig:tf_plot4_tgv}
\vspace{-.15in}
\end{figure*}

\begin{figure*} [!h]
\centering
\subfigure[{$u$ Channel.}]{ \label{fig:a}{}
\includegraphics[width=0.3\linewidth]{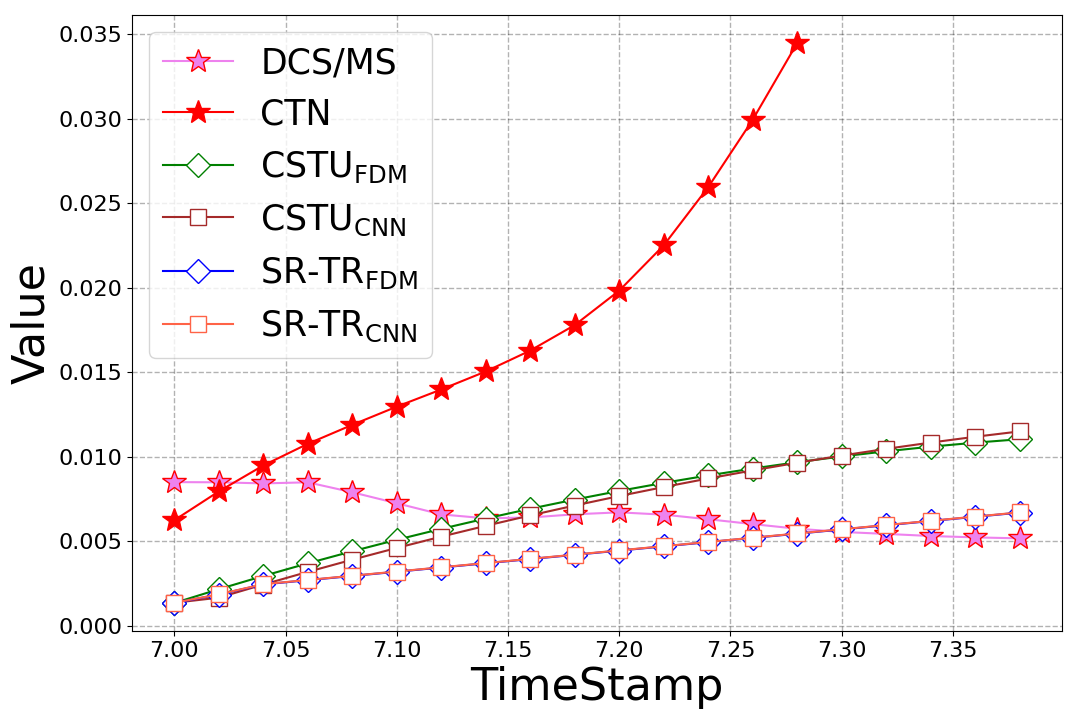}
}
\subfigure[{$v$ Channel.}]{ \label{fig:b}{}
\includegraphics[width=0.3\linewidth]{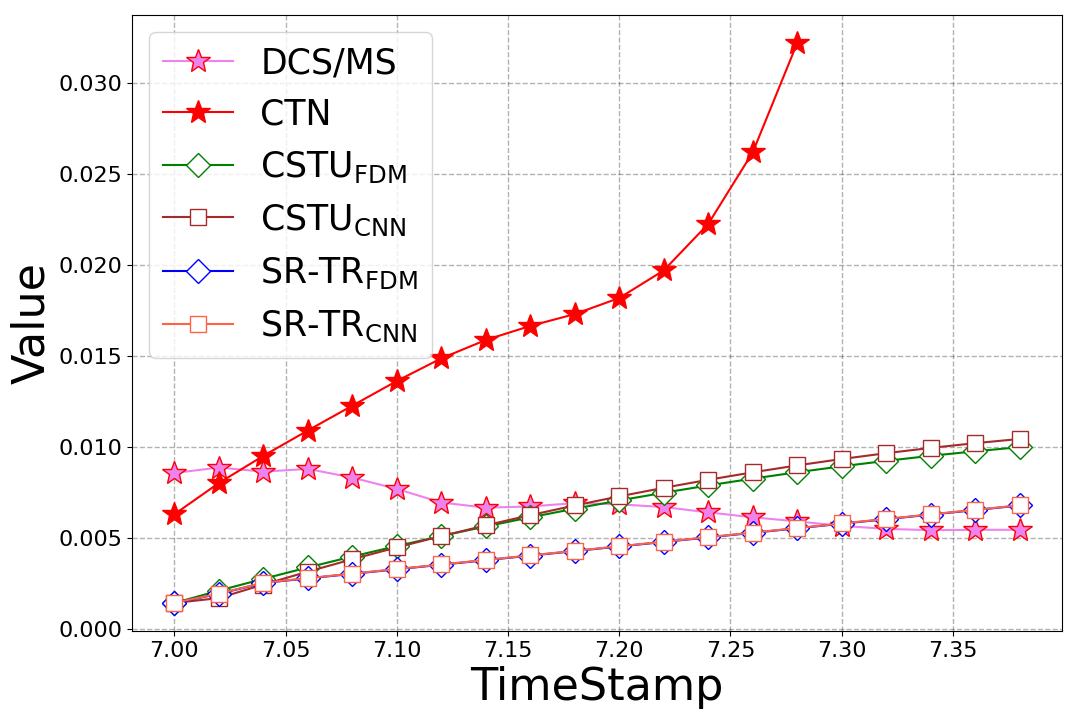}
}
\subfigure[{$w$ Channel.}]{ \label{fig:b}{}
\includegraphics[width=0.3\linewidth]{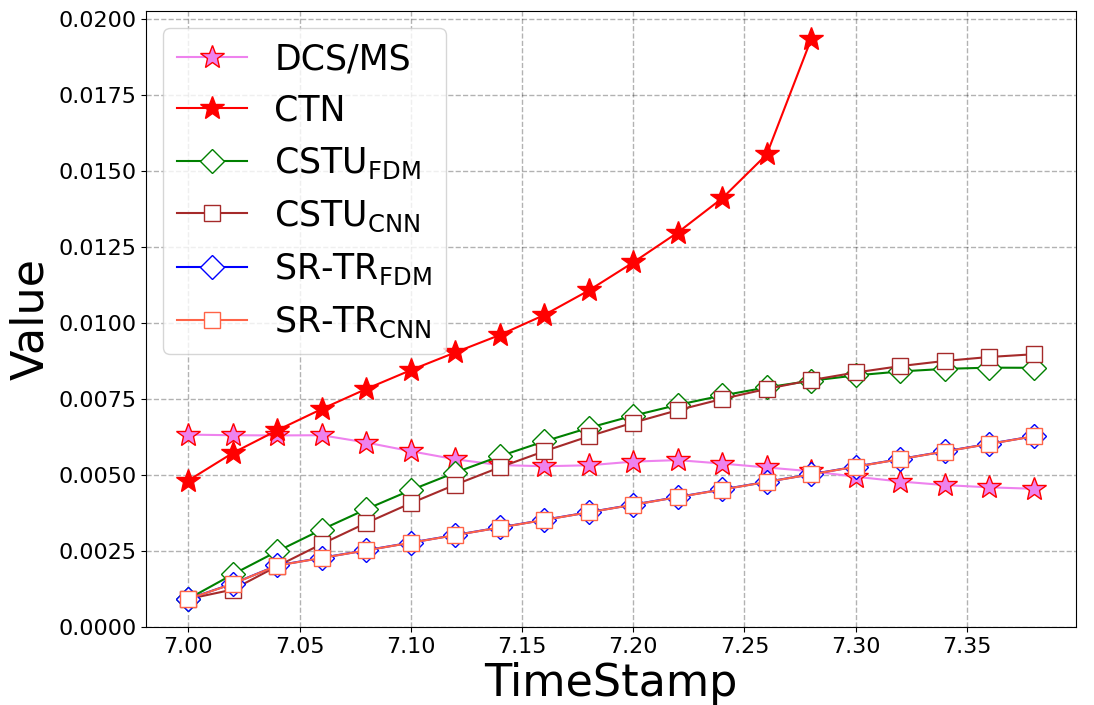}
}
\vspace{-.1in}
\caption{Change of dissipation difference by different models from 1st (80s) to 20th (120s) time step in TGV dataset.}
\label{fig:tf_plot5_tgv}
\vspace{-.1in}
\end{figure*}

\begin{figure*} [!h]
\centering
\subfigure[LES Upscaling.]{ \label{fig:a}
\includegraphics[width=0.15\linewidth]{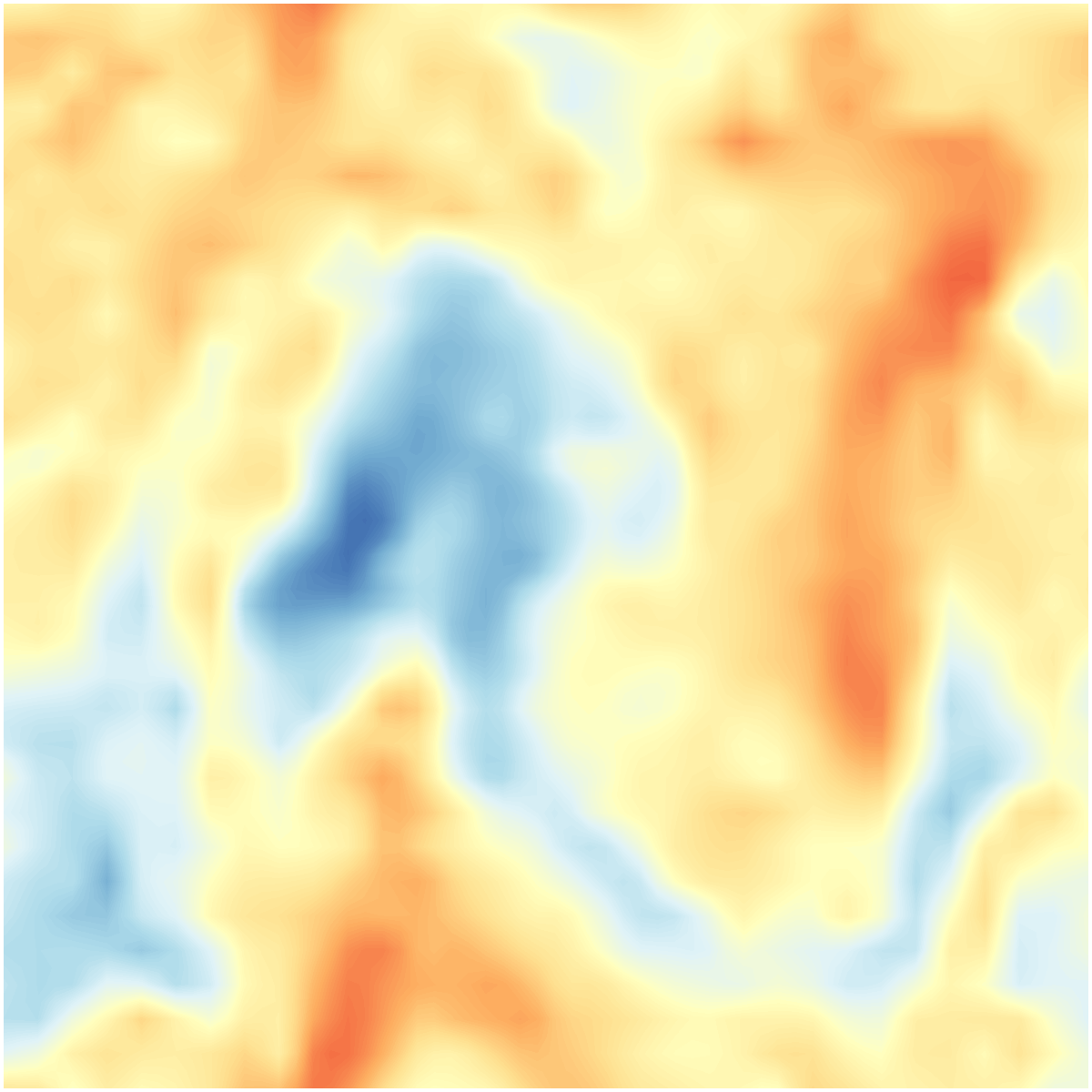}
}
\subfigure[DCS/MS.]{ \label{fig:b}
\includegraphics[width=0.15\linewidth]{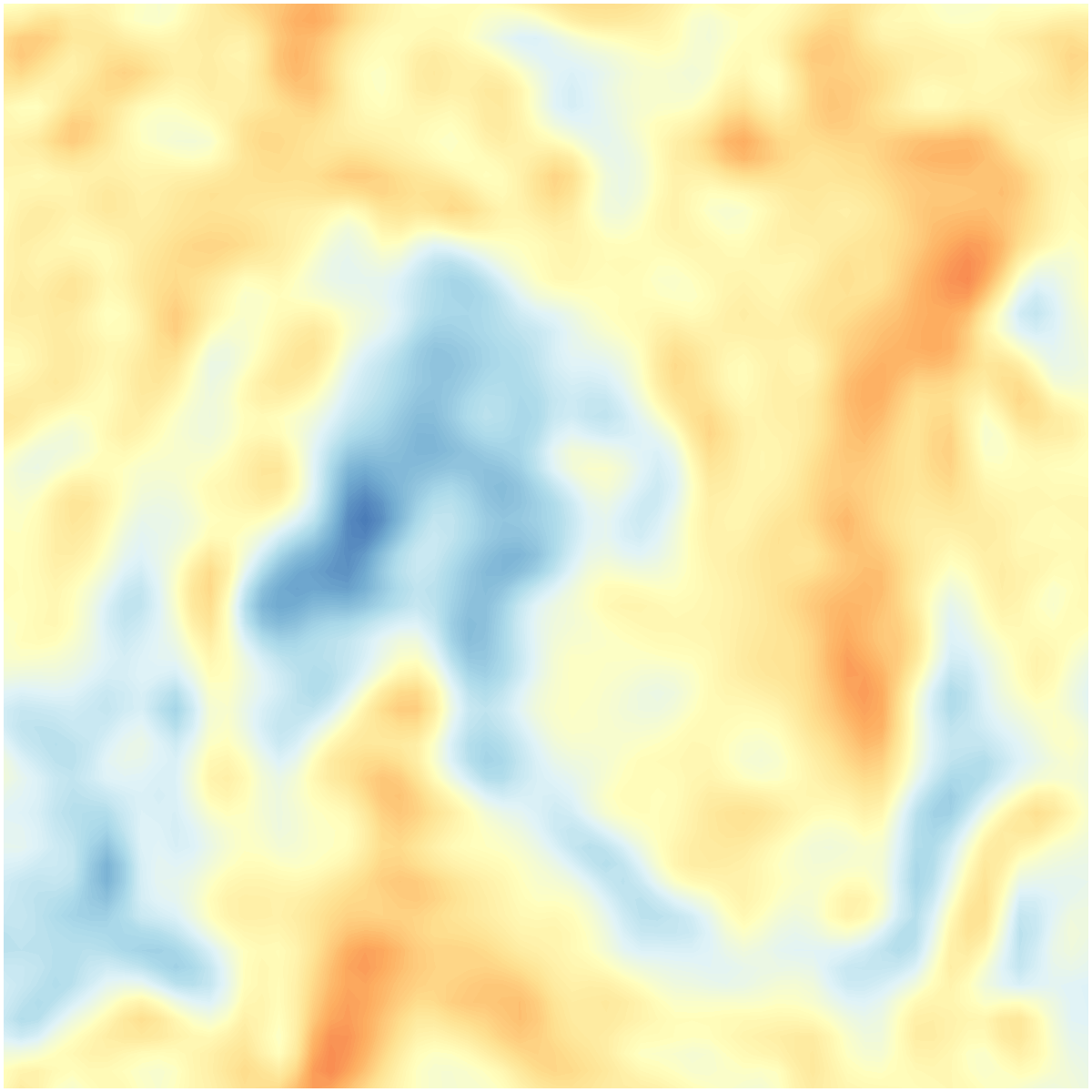}
}
\subfigure[CTN.]{ \label{fig:c}
\includegraphics[width=0.15\linewidth]{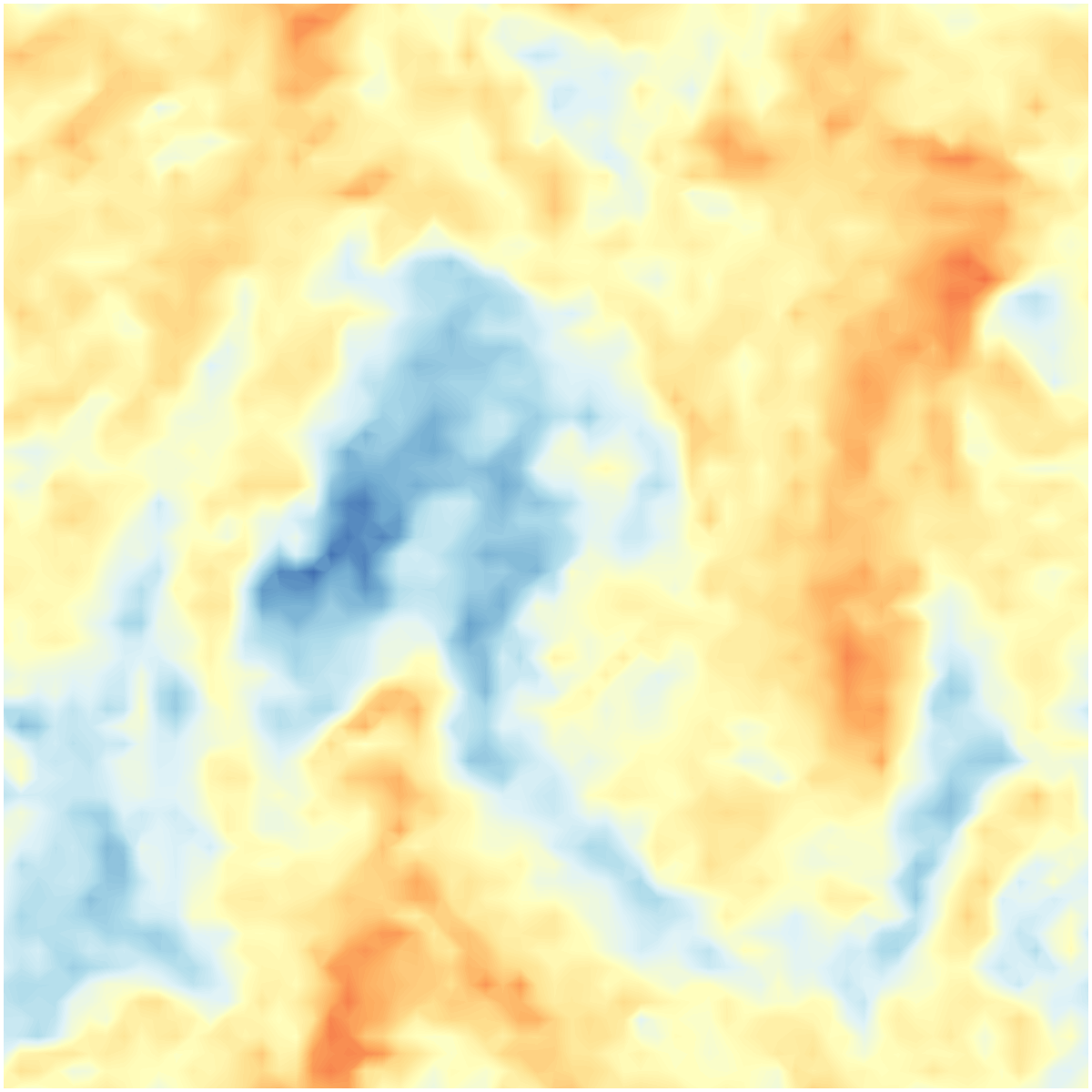}
}
\subfigure[$\text{SR-TR}_\text{FDM}$.]{ \label{fig:d}
\includegraphics[width=0.15\linewidth]{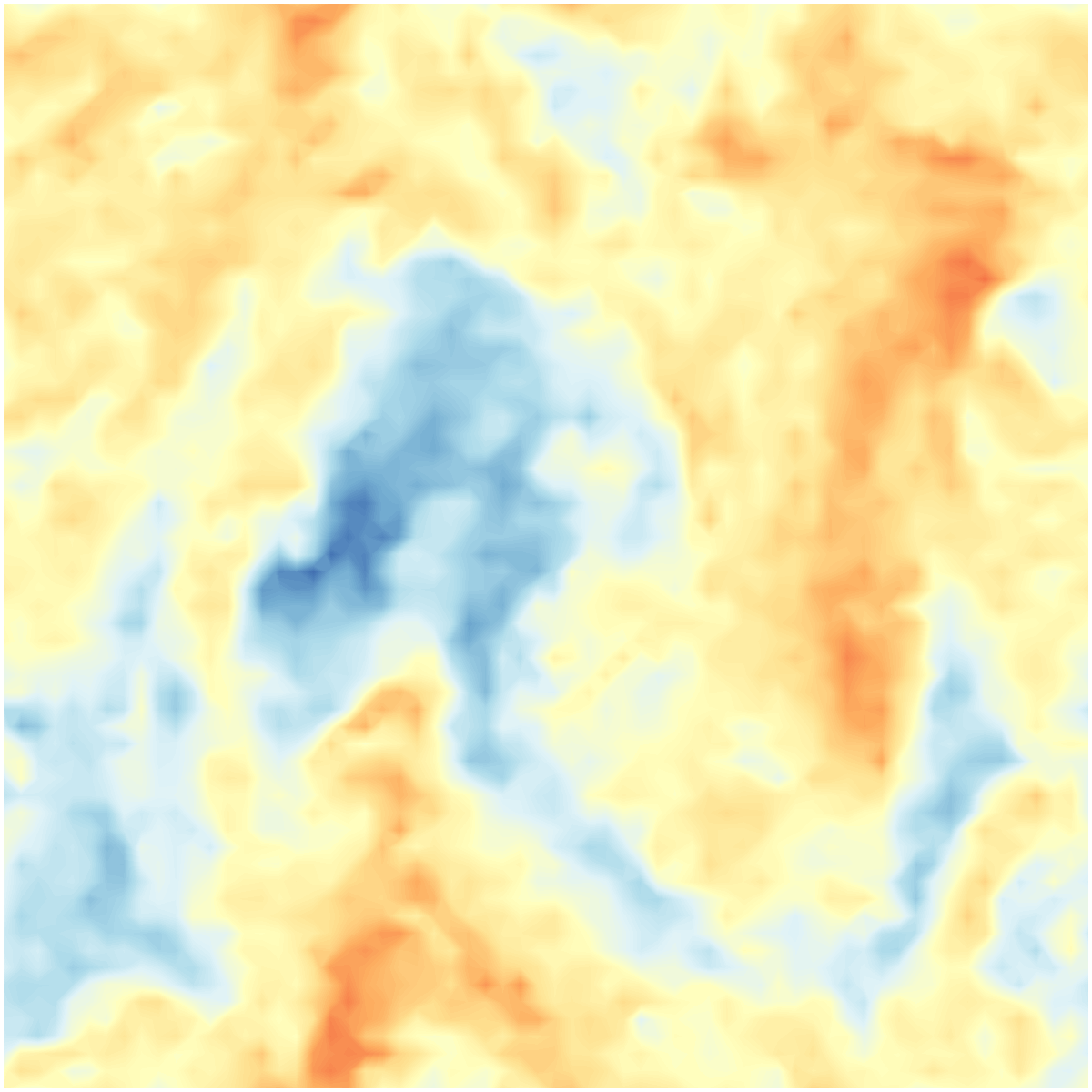}
}
\subfigure[$\text{SR-TR}_\text{CNN}$.]{ \label{fig:e}
\includegraphics[width=0.15\linewidth]{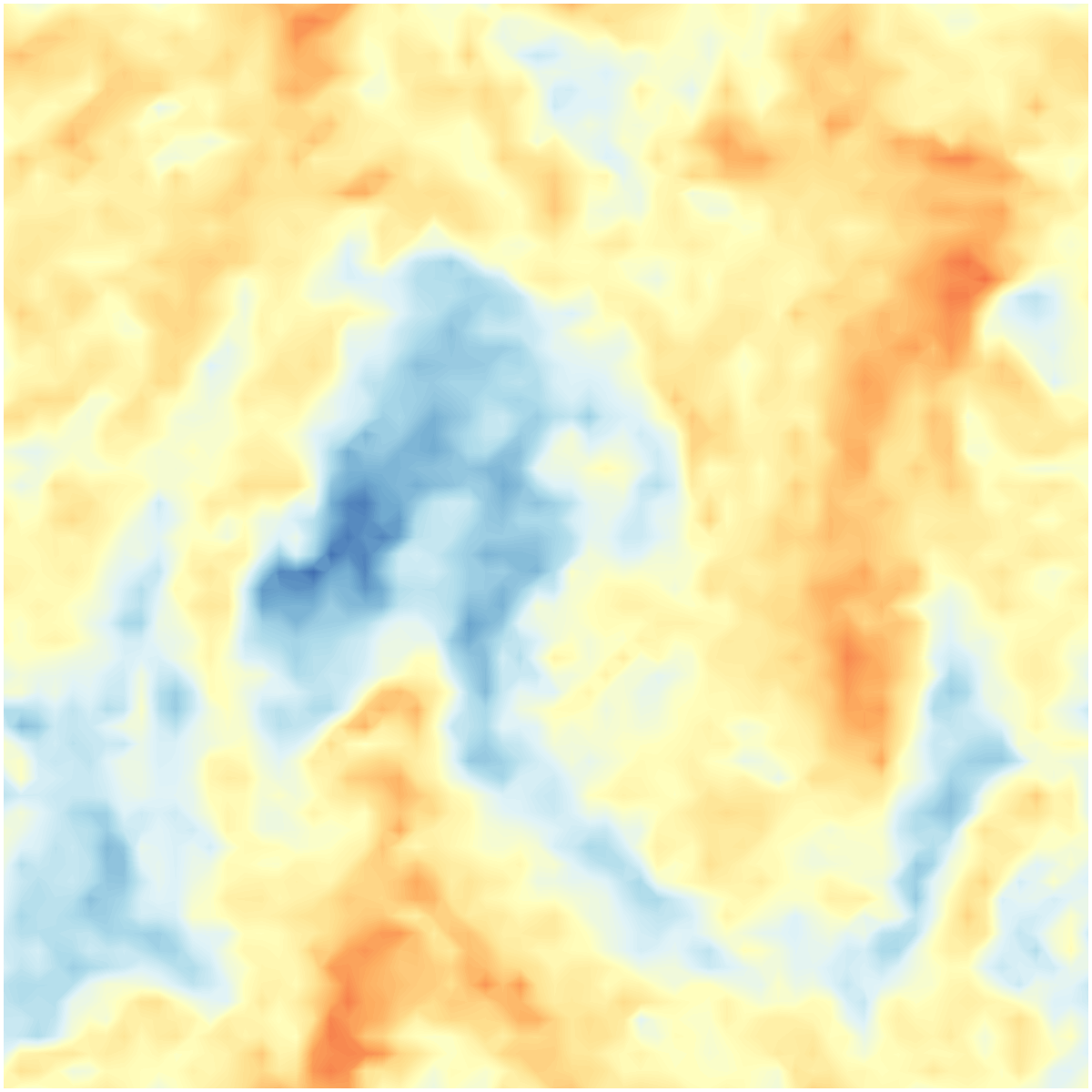}
}
\subfigure[Target DNS.]{ \label{fig:f}
\includegraphics[width=0.15\linewidth]{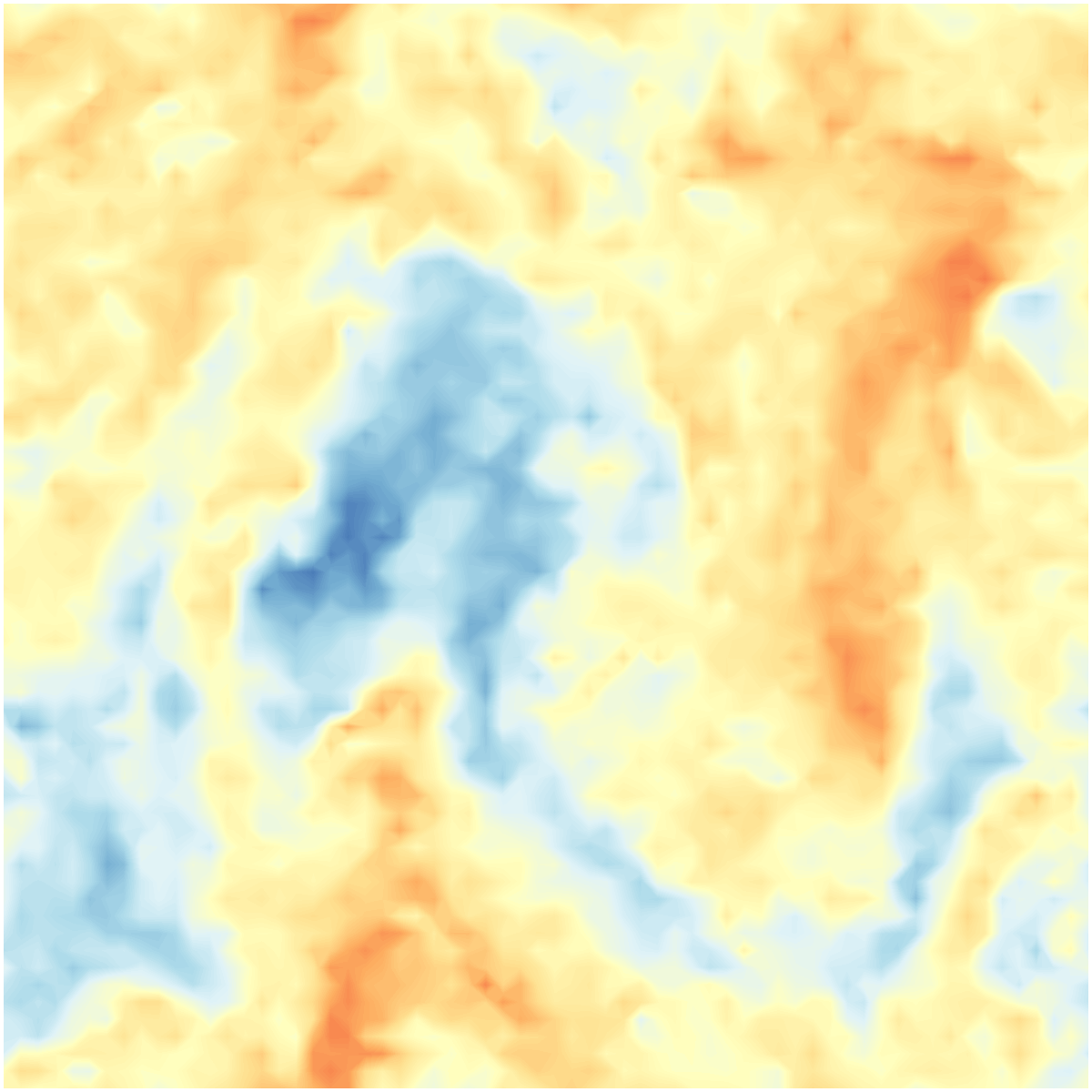}
}\vspace{-.1in}
\subfigure[LES Upscaling.]{ \label{fig:a}
\includegraphics[width=0.15\linewidth]{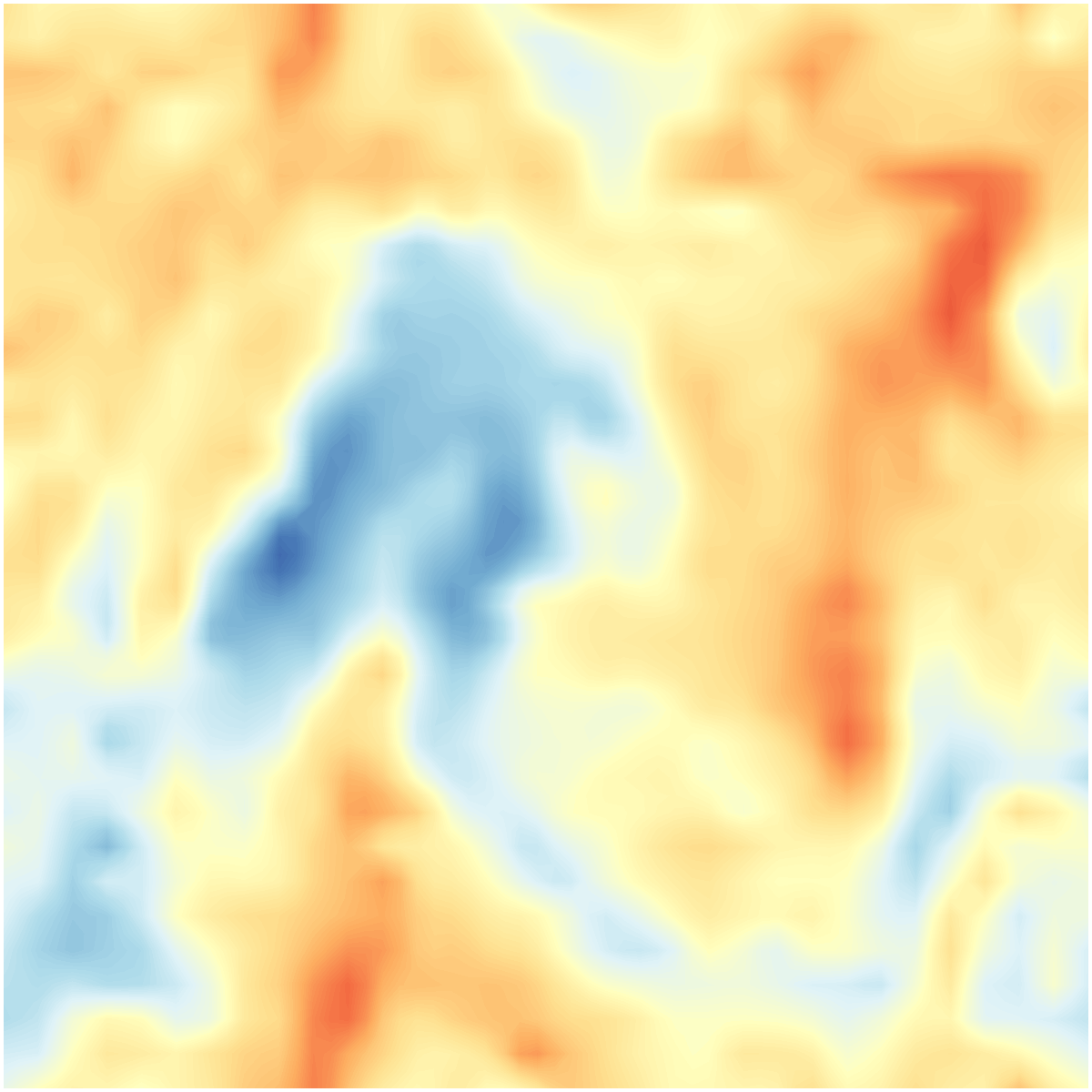}
}
\subfigure[DCS/MS.]{ \label{fig:b}
\includegraphics[width=0.15\linewidth]{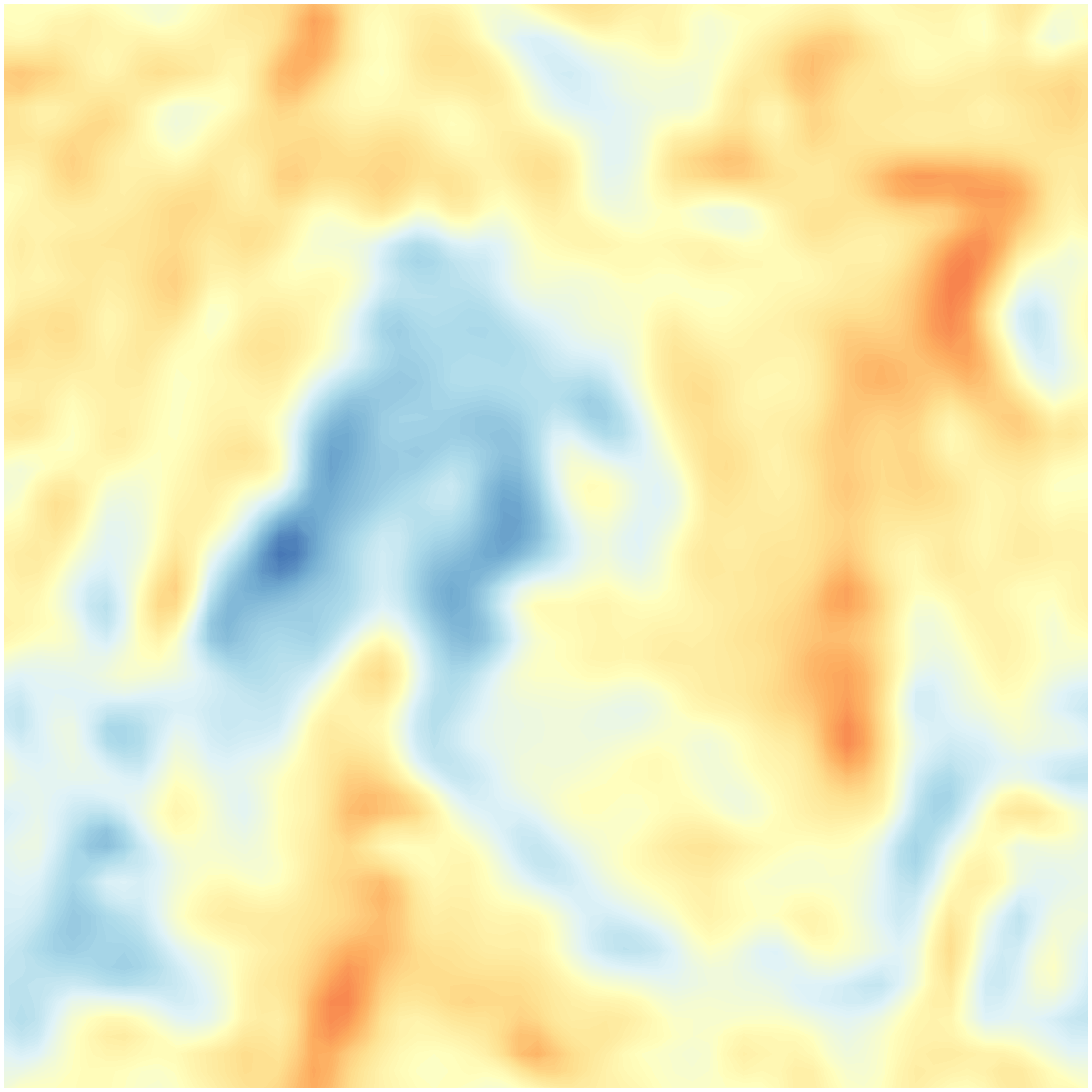}
}
\subfigure[CTN.]{ \label{fig:c}
\includegraphics[width=0.15\linewidth]{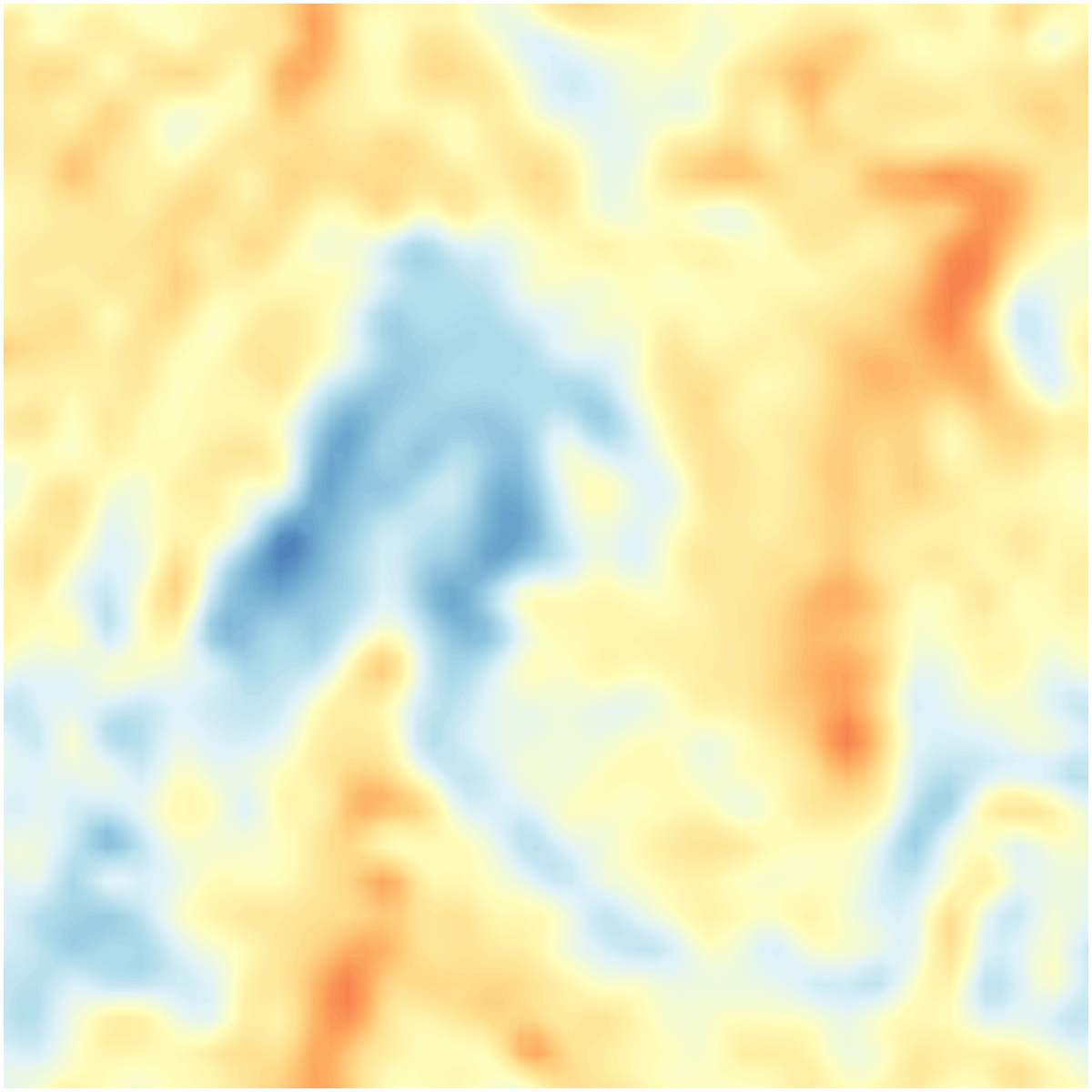}
}
\subfigure[$\text{SR-TR}_\text{FDM}$.]{ \label{fig:d}
\includegraphics[width=0.15\linewidth]{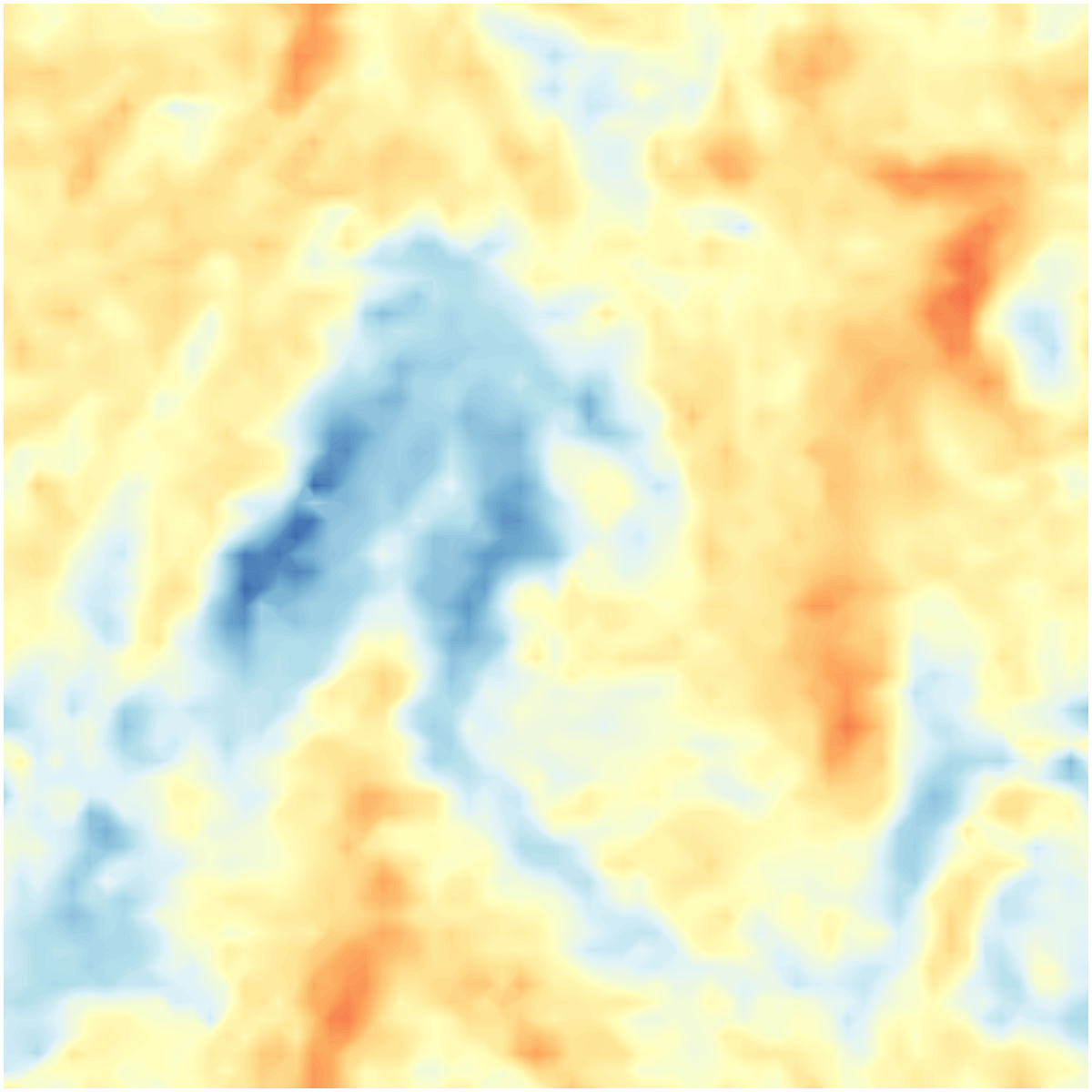}
}
\subfigure[$\text{SR-TR}_\text{CNN}$.]{ \label{fig:e}
\includegraphics[width=0.15\linewidth]{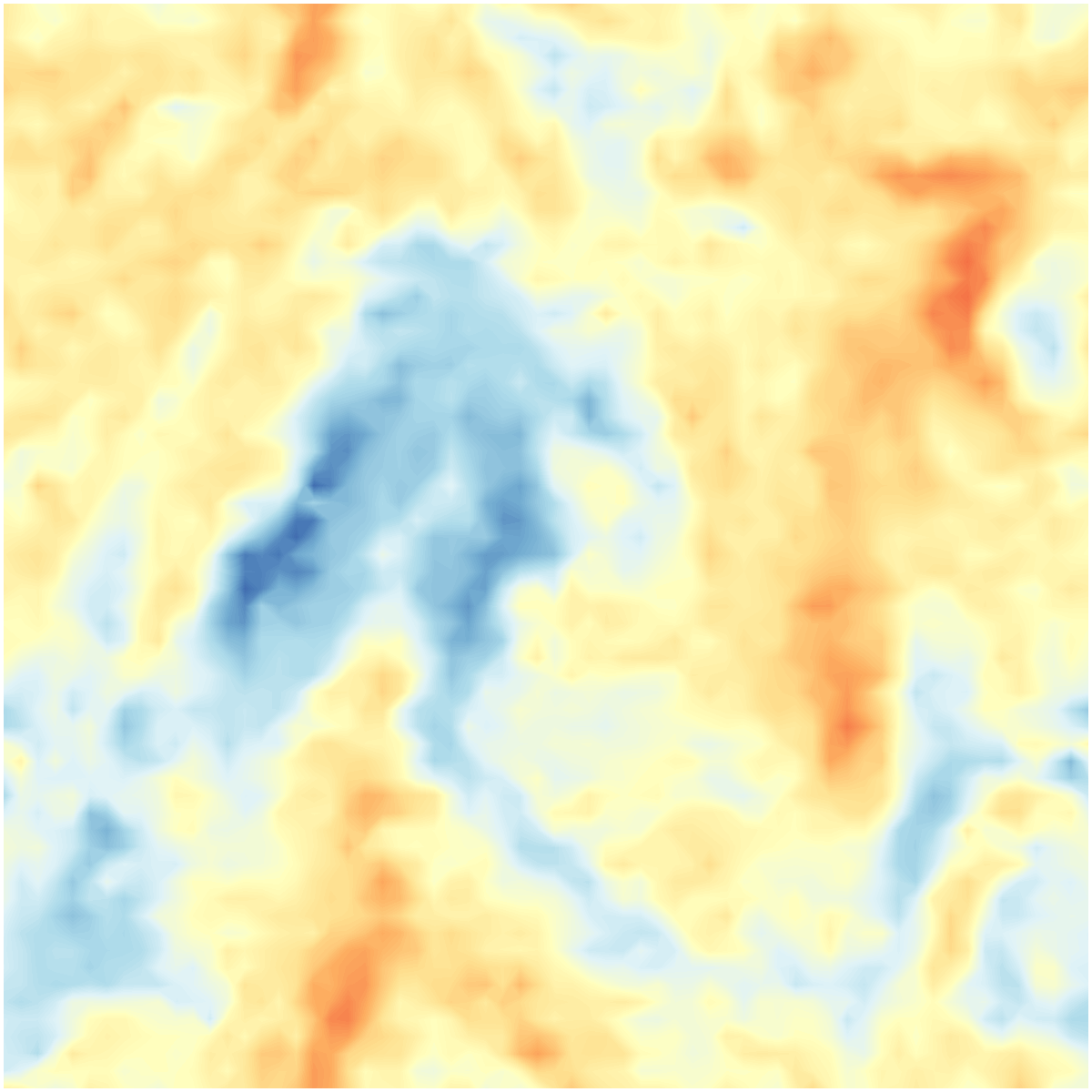}
}
\subfigure[Target DNS.]{ \label{fig:f}
\includegraphics[width=0.15\linewidth]{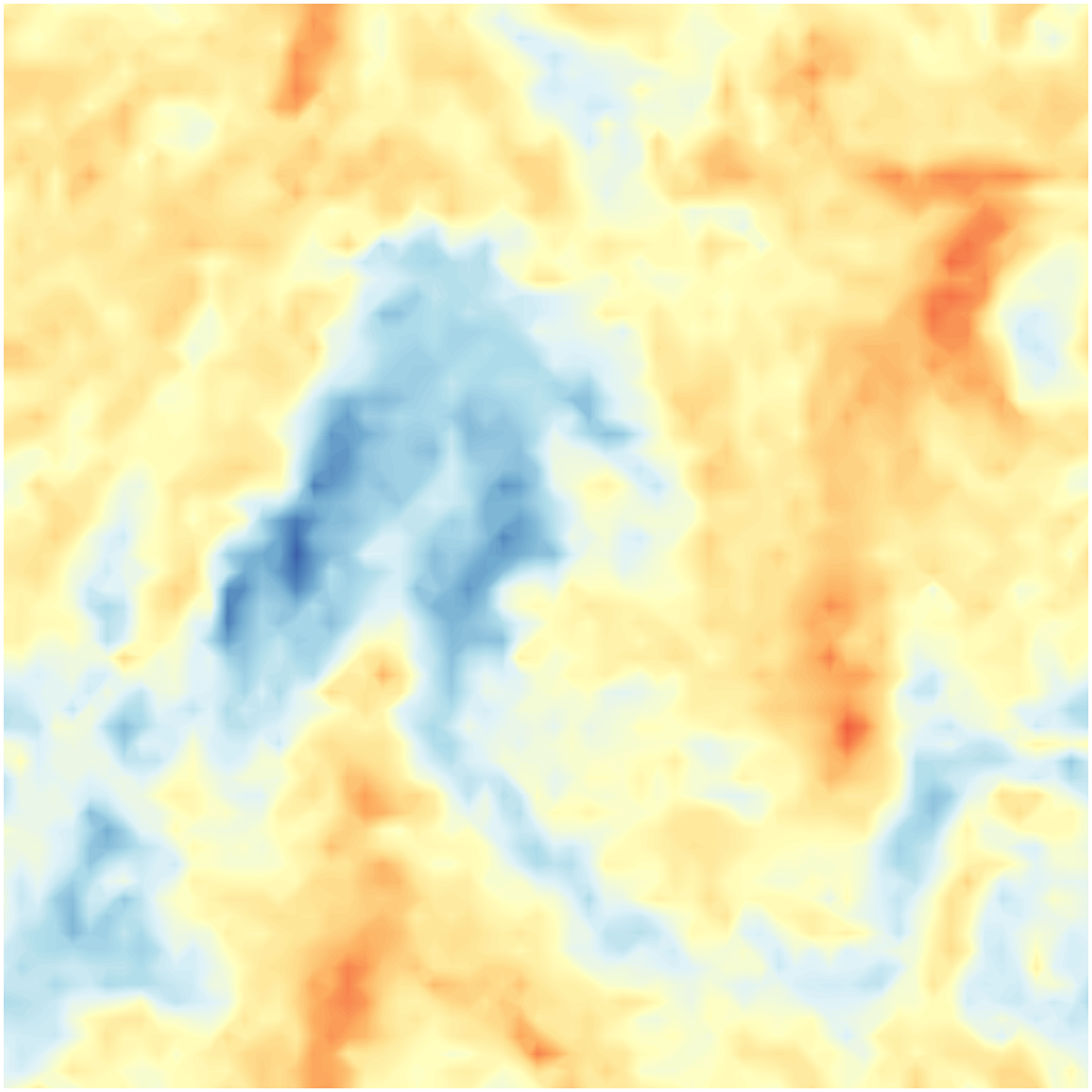}
}\vspace{-.1in}
\subfigure[LES Upscaling.]{ \label{fig:a}
\includegraphics[width=0.15\linewidth]{JHU/LES/w19.png}
}
\subfigure[DCS/MS.]{ \label{fig:b}
\includegraphics[width=0.15\linewidth]{JHU/DCS/w19.png}
}
\subfigure[CTN.]{ \label{fig:c}
\includegraphics[width=0.15\linewidth]{JHU/CTN/w19.png}
}
\subfigure[$\text{SR-TR}_\text{FDM}$.]{ \label{fig:d}
\includegraphics[width=0.15\linewidth]{JHU/SR_TR_FDM/w19.png}
}
\subfigure[$\text{SR-TR}_\text{CNN}$.]{ \label{fig:e}
\includegraphics[width=0.15\linewidth]{JHU/SR_TR_CNN/w19.png}
}
\subfigure[Target DNS.]{ \label{fig:f}
\includegraphics[width=0.15\linewidth]{JHU/GT/w19.png}
}
\vspace{-.1in}
\caption{Reconstructed $w$ channel by each method on a sample testing slice along the $z$ dimension in the FIT dataset. The reconstruction results are shown at 1st (5.6s), 10th (5.8s) and 20th (6s) in (a)-(f), (g)-(l), and (m)-(r), respectively.}
\label{fig:tf_plot3}
\end{figure*}

\begin{figure} [!t] 
\centering
\includegraphics[width=0.7\columnwidth]{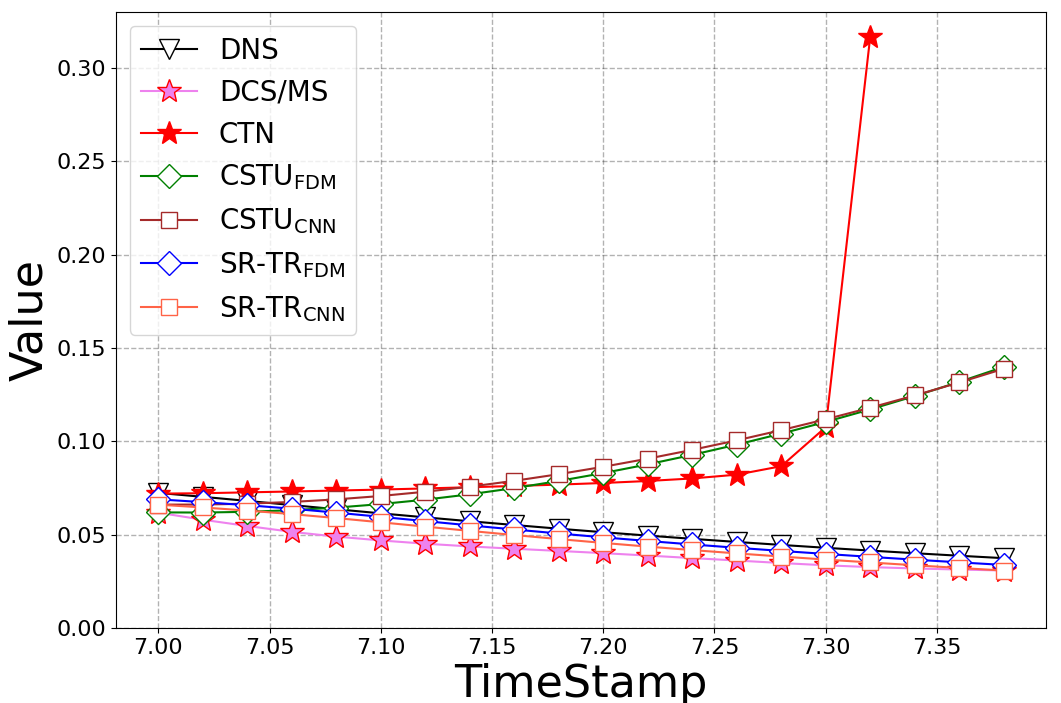}
\vspace{-0.05in}
\caption{Change of kinetic energy produced by the reference DNS and different models in the TGV datasets.}
\label{fig:kinetic1}
\end{figure}

\begin{figure*} [!h]
\centering
\subfigure[LES Upscaling.]{ \label{fig:a}
\includegraphics[width=0.15\linewidth]{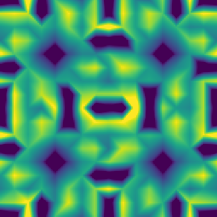}
}
\subfigure[DCS/MS.]{ \label{fig:b}
\includegraphics[width=0.15\linewidth]{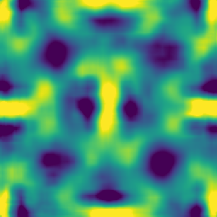}
}
\subfigure[CTN.]{ \label{fig:c}
\includegraphics[width=0.15\linewidth]{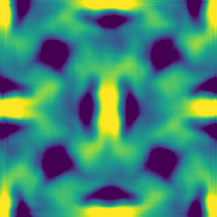}
}
\subfigure[$\text{SR-TR}_\text{FDM}$.]{ \label{fig:d}
\includegraphics[width=0.15\linewidth]{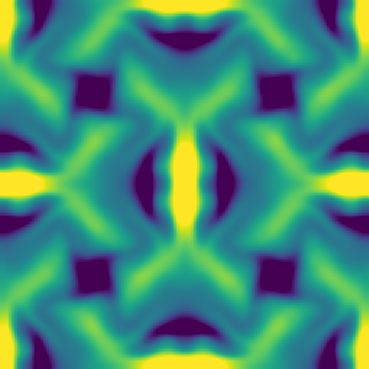}
}
\subfigure[$\text{SR-TR}_\text{CNN}$.]{ \label{fig:e}
\includegraphics[width=0.15\linewidth]{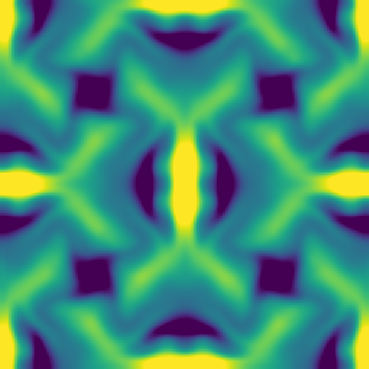}
}
\subfigure[Target DNS.]{ \label{fig:f}
\includegraphics[width=0.15\linewidth]{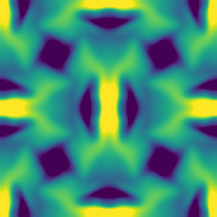}
}\vspace{-.1in}
\subfigure[LES Upscaling.]{ \label{fig:a}
\includegraphics[width=0.15\linewidth]{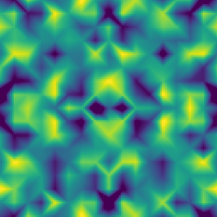}
}
\subfigure[DCS/MS.]{ \label{fig:b}
\includegraphics[width=0.15\linewidth]{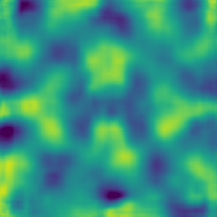}
}
\subfigure[CTN.]{ \label{fig:c}
\includegraphics[width=0.15\linewidth]{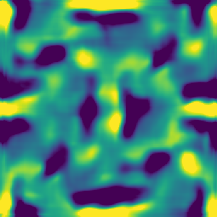}
}
\subfigure[$\text{SR-TR}_\text{FDM}$.]{ \label{fig:d}
\includegraphics[width=0.15\linewidth]{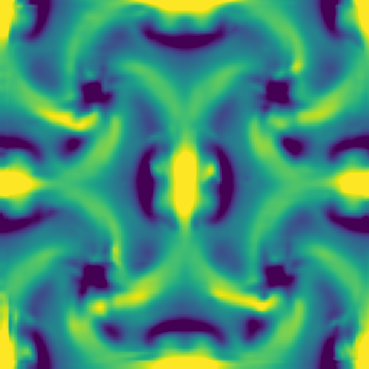}
}
\subfigure[$\text{SR-TR}_\text{CNN}$.]{ \label{fig:e}
\includegraphics[width=0.15\linewidth]{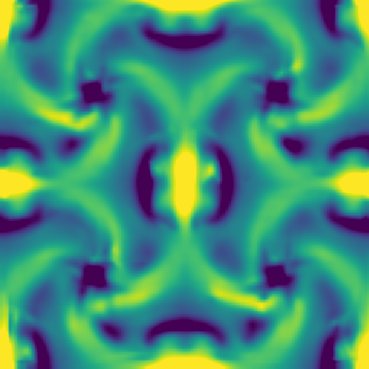}
}
\subfigure[Target DNS.]{ \label{fig:f}
\includegraphics[width=0.15\linewidth]{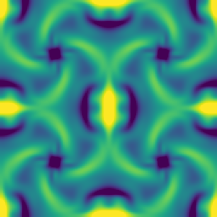}
}\vspace{-.1in}
\subfigure[LES Upscaling.]{ \label{fig:a}
\includegraphics[width=0.15\linewidth]{TGV/LES/w_14.png}
}
\subfigure[DCS/MS.]{ \label{fig:b}
\includegraphics[width=0.15\linewidth]{TGV/DCS/w_14.png}
}
\subfigure[CTN.]{ \label{fig:c}
\includegraphics[width=0.15\linewidth]{TGV/CTN/w_14.png}
}
\subfigure[$\text{SR-TR}_\text{FDM}$.]{ \label{fig:d}
\includegraphics[width=0.15\linewidth]{TGV/SR_TR_FDM/w14.png}
}
\subfigure[$\text{SR-TR}_\text{CNN}$.]{ \label{fig:e}
\includegraphics[width=0.15\linewidth]{TGV/SR_TR_CNN/w14.png}
}
\subfigure[Target DNS.]{ \label{fig:f}
\includegraphics[width=0.15\linewidth]{TGV/GT/w_14.png}
}
\vspace{-.1in}
\caption{Reconstructed $w$ channel by each method on a sample testing slice along the $z$ dimension in the TGV dataset. The reconstruction results are shown at 1st (80s), 10th (100s), and 15th (110s) in (a)-(f), (g)-(l), and (m)-(r), respectively.}
\label{fig:tf_plot4}
\end{figure*}

\section{Experimental Settings}
\subsection{Dataset}
\label{sec:dataset}
To evaluate the performance of the proposed methodology, the data sets pertaining to two turbulent flows are considered: a forced isotropic turbulent flow (FIT)~\cite{FITdata} and the Taylor-Green vortex (TGV)~\cite{brachet1984taylor} flow. In both scenarios, the mean velocity is zero, $\overline{\bf Q}(t)=0$, and the Reynolds number is sufficiently high to induce turbulent characteristics in the flow.

The FIT dataset comprises the original DNS records of forced isotropic turbulence, representing an incompressible flow. 
The flow is subjected to energy injection at low wave numbers as part of the forcing mechanism. The DNS data consists of $5024$ time steps, with each step separated by a time interval of $0.002$s, encompassing both velocity and pressure fields. For this study, the original DNS data is generated to three different grids: $128 \times 64 \times 64$, $128 \times 128 \times 128$, and $128 \times 256 \times 256$. Simultaneously, the LES data is generated on grids of size $128 \times 32 \times 32$. Both DNS and LES data are collected along the $128$ equally spaced grid points along the $z$ axis.

The Taylor-Green vortex (TGV) represents another incompressible flow. The evolution of the TGV involves the elongation of vorticity, resulting in the generation of small-scale, dissipating eddies. A box flow scenario is examined within a cubic periodic domain spanning $[-\pi,\pi]$ in all three directions. The initial conditions are defined as:
\begin{equation}
\begin{aligned}
u (x,y,z,0) &= \sin(x) \cos(y) \cos(z),\\ 
v(x,y,z,0) &= - \cos(x)\sin(y)\cos(z),\\ 
w(x,y,z,0) &= 0.   
\end{aligned}
\end{equation}
The DNS and LES resolutions are  $ 128 \times 128 \times 65  $ and $ 32 \times 32 \times 65$, respectively. Both DNS and LES data are produced along the $65$ equally-spaced grid points along the $z$ axis.

\subsection{SR-TR method and baselines} The performance of the SR-TR method is evaluated and compared with multiple existing methods for SR and turbulent flow downscaling. Specifically, we implement the SR-TR-based methods: $\text{ST-TR}_\text{FDM}$ and $\text{ST-TR}_\text{CNN}$ using FDM and CNN to approximate spatial gradients, respectively. Additionally, four popular SR methods SRCNN~\cite{dong2014learning}, RCAN~\cite{zhang2018image}, HDRN~\cite{Duong2021}, and SRGAN~\cite{ledig2017photo}, two well-known dynamic fluid downscaling methods: DCS/MS~\cite{fukami2019super} and FSR~\cite{yang2023super}, and the Fourier neural operator (FNO)~\cite{li2020fourier}, are used as baselines. 

To better verify the effectiveness of each of the model's components,  three additional baselines are also introduced: convolutional transition network (CTN), $\text{CSTU}_\text{FDM}$, and $\text{CSTU}_\text{CNN}$~\cite{bao2022physics}. The CTN is created by combining SRCNN and LSTM~\cite{LSTM}. $\text{CSTU}_\text{FDM}$ and $\text{CSTU}_\text{CNN}$ are similar to $\text{ST-TR}_\text{FDM}$ and $\text{ST-TR}_\text{CNN}$, but they are created without using the degradation-based refinement.  The objective of comparing the CTN with CSTU-based methods is to demonstrate the advantages of  CSTU in spatio-temporal DNS reconstruction. The advantages of the refinement process are demonstrated by comparing CSTU-based and  SR-TR-based methods.
Furthermore, to present additional evidence regarding the potential enhancement of flow data reconstruction through additional LES input, 
we  implement a variant of the proposed method $\text{ST-TR}_\text{FDM}^\text{a}$, which utilizes LES data as additional input.  

\subsection{Implementation details} The proposed SR-TR method is implemented via Tensorflow 2 with an A100 GPU. The model is first trained in 500 epochs with ADAM optimizer~\cite{kingma2014adam_arxiv} from an initial learning rate of $0.0001$. In the refinement step,  the learning rate is lowered to $0.00005$, and the training rate is iterated by 10 epochs. All the hidden variables and gating variables are in $32$ dimensions. The values of $\alpha_0$, $\alpha_1$, and $\alpha_2$ are set as  $1000,\ 1,$ and $1$, respectively.

\subsection{Evaluation Metrics}
The assessment of DNS reconstruction performance employs two metrics: the Structural Similarity Index Measure (SSIM)~\cite{wang2004image} and dissipation difference ~\cite{enwiki:1127277109}. SSIM measures the similarity between reconstructed and target DNS data in terms of luminance, contrast, and overall structure. Higher SSIM values indicate better reconstruction. Dissipation evaluates the model's gradient capturing ability, considering dissipation for each velocity vector component ($u$, $v$, and $w$). The dissipation operator is defined by:

\begin{equation}
\small
\chi (Q) \equiv \nabla Q \cdot \nabla Q= \left(\frac{\partial Q}{\partial x}\right)^2 + \left(\frac{\partial Q}{\partial y}\right)^2 + \left(\frac{\partial Q}{\partial z}\right)^2.
\end{equation}
The dissipation is used to measure the difference in flow gradient between the true DNS and generated data. This is  represented by $|\chi({Q}^d) - \chi(\hat{{Q}}^d)|$. The lower value of this difference indicates better performance.  

\section{Reconstruction Performance}
\subsection{Temporal analysis} 
For the FIT dataset, the performance for reconstructing DNS is measured for each step during a $0.4s$ period ($20$ time steps) in the testing phase. The performance change using SSIM and dissipation difference are shown in  Fig.~\ref{fig:tf_plot1} and Fig.~\ref{fig:tf_plot2}. Several observations are highlighted: (1)~With larger time intervals between training and prediction data, the performance becomes worse. In general, SR-TR-based methods exhibit greater stability over long-term prediction, indicating superior performance compared to other methods. (2)~SR-TR-based methods outperform CSTU-based methods, illustrating the effectiveness of degradation-based refinement in mitigating prediction bias over long-term predictions. (3)~Both $\text{CSTU}_\text{FDM}$ and $\text{CSTU}_\text{CNN}$ demonstrate similar performance. A parallel observation arises when comparing the two variants of SR-TR-based methods, demonstrating that either approach for estimating the spatial derivative within the CSTU can obtain similar performance. The same conclusion can be drawn from the temporal analysis of the TGV data, shown in Figs.~\ref{fig:tf_plot4_tgv} and ~\ref{fig:tf_plot5_tgv}. 

\subsubsection{Visualization.} In Figs.~\ref{fig:tf_plot3}, the reconstructed data are presented for multiple time steps (1st, 10th, and 20th) after the training phase. These figures depict slices of the $w$ component at specific $z$ values. During the 1st step, both the SR-TR-based methods and the baseline CTN model show highly accurate reconstruction results. This outcome arises due to the similarity between the test data and the training data from the last time step. Meanwhile, the baseline DSC/MS exhibits diminishing performance starting from the early times. As time progresses, the SR-TR-based methods consistently outperform the baseline methods. Especially, a significant divergence emerges at the 20th time step. Most of the baseline methods struggle to capture the correct flow transport pattern, whereas the SR-TR-based methods indicate considerably improved performance in the later stages. Similar trends can be observed in the TGV dataset, presented in Figs.~\ref{fig:tf_plot4}.

\subsubsection{Validation via physical metrics.} 
The performance is also evaluated through the long-term prediction of turbulent kinetic energy. Figure~\ref{fig:kinetic1} illustrates the energies corresponding to the target DNS, along with the reconstructed flow data from both the baselines and SR-TR-based methods for the FIT dataset. The following observations can be made: (1)~The SR-TR-based methods exhibit superior performance compared to the baseline methods DCS/MS and CTN. (2)~The performance of the CSTU-based methods degrades significantly after the 10th time step. This amplification of accumulated errors at each time step contributes to this outcome. A similar conclusion can be drawn from the analysis of TGV data in supplementary file. (3) CTN almost fails to capture the kinetic energy after the 10th time step, leading to an explosion.

\end{document}